\documentclass{jfm}
\usepackage[T2A]{fontenc} % Russian 1
\usepackage[utf8]{inputenc} % Russian 2
\usepackage[russian,english]{babel} % Russian 3 This can be exchanged with the [russian,english]
\usepackage{graphicx}
\usepackage{epstopdf, epsfig}
\usepackage{color}
\usepackage{ulem}
\newcommand{\ba}{\begin{array}}
\newcommand{\ea}{\end{array}}
\newcommand{\be}{\begin{equation}}
\newcommand{\ee}{\end{equation}}
\newcommand{\bea}{\begin{eqnarray}}
\newcommand{\eea}{\end{eqnarray}}
\newcommand{\bfig}{\begin{figure}}
\newcommand{\efig}{\end{figure}}
\newcommand{\Bl}{\Bigl}
\newcommand{\Br}{\Bigr}

\newcommand{\re}{{\rm e}}

\newcommand{\pl}{\partial}
\newcommand{\dd}{{\rm d}}
\newcommand{\al}{\alpha}

\newcommand{\vp}{\varphi}
\newcommand{\vep}{\varepsilon}

\newcommand{\din}{\displaystyle\int\limits}

\newcommand{\II}{\displaystyle\int\limits^{\infty}_}

\newcommand{\dfrac}{\displaystyle\frac}
\shorttitle{Reflectionless wave propagation}
\shortauthor{S. M. Churilov and Y. A. Stepanyants}

\title{Reflectionless wave propagation on shallow water with variable bathymetry and current}

\author{Semyon M. Churilov\aff{1}
     \and Yury A. Stepanyants\aff{2,}\aff{3}
\corresp{\email{Yury.Stepanyants@usq.edu.au}}}

\affiliation{\aff{1} Institute of Solar-Terrestrial Physics of the Siberian Branch of the Russian Academy of Sciences, PO Box 291, Irkutsk, 664033, Russia
\aff{2} Scool of Sciences, University of Southern Queensland, West St., Toowoomba, QLD, 4350, Australia \\
\aff{3} Department of Applied Mathematics, Nizhny Novgorod State Technical University \\n.a. R. E. Alekseev, 24 Minin St., Nizhny Novgorod, 603950, Russia}

\begin{document}

\maketitle

\begin{abstract}
In the linear approximation, we study a one-dimensional problem of the reflectionless wave propagation on a surface of a shallow duct with the spatially varying water depth, duct width, and current. We show that both global and bounded exact solutions describing reflectionless propagation in opposite directions of long waves of arbitrary shape exist for the particular variations of these parameters. A general analysis of the problem is illustrated by a few solutions constructed for the specific cases of spatial profiles of the flow velocity. The results obtained can be of interest to mitigate the possible impact of waves on ships, marine engineering constructions, and human activity in the coastal zones.
\end{abstract}


\section{Introduction}
\label{Sect01}

Wave propagation in an inhomogeneous ocean is one of the important and topical problems of physical oceanography and fluid mechanics, in general. 
However, it is a difficult problem from the mathematical point of view in the general statement due to the nonlinearity, dispersion, and strong inhomogeneity. 
Analytical solutions either in the exact or approximate forms are possible in exceptional cases; see, for example, \citep{Stoker-57, Sretensky-77} in application to water waves. 
In these books, exact solutions for small-amplitude surface waves on water of arbitrary depth with the sloping bottom are presented. 
When the characteristic length of a small-amplitude wave is short in comparison with the typical scale of inhomogeneity, the well-known WKB approximation can be used for the description of quasi-monochromatic waves.
However, in some special cases of media parameters variation, exact solutions can be derived for waves of arbitrary length both in the linear and even in the nonlinear cases. 
There is a vast volume of publications on this theme in the past two decades; here we refer only to the most recent and relevant, others can be found in the cited publications \citep{Did-Pel-Soom-09, Did-Pel-09, Didenkulova-09, Dobr-10, Dobr-11, Grim-10, Did-Pel-11, Dobr-13, Pelin-17, Pelinovsky-17, Pelinovsky-19}.
The outcomes of theoretical predictions for shallow-water waves were validated in the numerical modelling (see, for example, \citep{Choi-08, Vlasenko-87, Pudjaprasetya-21}). 
The important feature of such solutions is they describe the {\it reflectionless} wave propagation when the wave energy is transmitted through the inhomogeneous zone the most effectively without the energy losses on the wave reflection. 
Such situations can be the most dangerous from the point of view of their possible impact on ships and engineering constructions in the coastal zones.

The problem of reflectionless wave propagation in inhomogeneous media has a long history; it was studied for  monochromatic waves in plasma physics \citep{Ginzburg-70, Petrukhin-20}, acoustics \citep{Brekhovskikh-80}, solids \citep{Clements-74}, fluids \citep{Magaard-62, Vlasenko-87}, etc. 
A more general analysis of reflectionless propagation of water waves of different shapes was presented in the cited above papers by Didenkulova, Pelinovsky, Dobrokhotov et al. for the particular configurations of a bottom profile. 
In the cited publications, wave propagation was studied basically for the fluid without mean currents \citep[in the paper by][a particular case of a variable mean current was taken into account]{Dobr-13}. 
However, in many cases, influence of currents on wave propagation can be significant and therefore, should be studied from the general point of view. 

In this paper, we fill the gap and study in the long-wave approximation the reflectionless linear wave propagation in canals with the variable depth, width, and currents. 
The problem is formulated in terms of the velocity potential which allows us to find spatially varying mean flow profile, as well as variation of the duct width and depth, which admit the reflectionless wave propagation in ducts. 
We show that both the global solutions defined on the entire $x$-axis and bounded solutions defined only on the semi-infinite $x$-interval do exist and present particular examples. Then, we find the conditions which secure the existence of global solutions. 
The results obtained are illustrated graphically.

To solve this problem, we employ the transformation technique which has been used in many publications \citep[see, for example,][and references therein]{Grimshaw-10, Pelinovsky-17}. Such a technique is one of the particular cases of the general methods of reduction of linear differential equations to some reference equations which can be solved analytically. 
These methods develop starting back from the works of Euler, Laplace, Poisson, Darboux, et al., until now \citep[see, for example,][and references therein]{Bluman-83, Varley-88, Chirkunov-14, Kaptsov-21}.

The paper is organized as follows. In Section \ref{Sect02}, we formulate the basic equations and derive conditions under which the wave propagation is reflectionless. 
The Section ends with a simple condition that allows us to distinguish a particular class of global solutions.
Section \ref{Sect03} is devoted to the general analysis of the properties of solutions belonging to another class among which there are both global and bounded solutions. 
Particular solutions of this class are obtained and analyzed in Section \ref{Sect04}. 
In Section \ref{Sect05}, we explore the existence of smooth global solutions belonging to the second class and derive sufficient conditions of their existence. 
The possibility to construct global solutions by matching bounded solutions is considered in Section \ref{Sect06}. 
Finally, Section \ref{Sect07} contains a discussion of results and concluding remarks. 
The technical aspects of matching solutions are considered in Appendix.
\section{Problem statement and general analysis}
\label{Sect02}

\begin{figure}
%\vspace{-3.5cm}
\centerline{\includegraphics[width=0.8\textwidth]{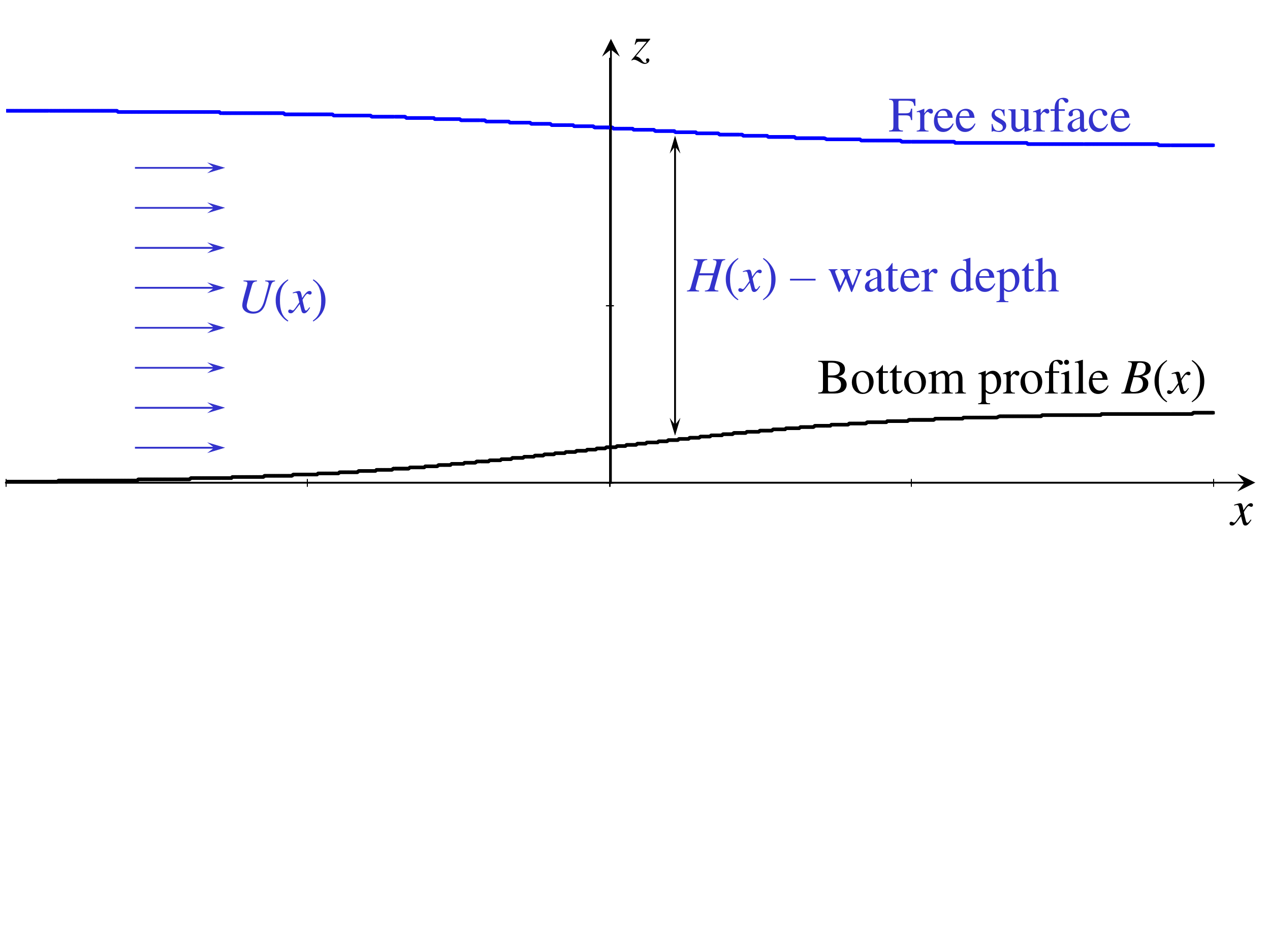}}%
\vspace{-3.5cm}
\caption{(Color online.) Sketch of the flow configuration in the vertical plane.}
\label{f00}
\end{figure}

Let us consider the propagation of surface waves on a shallow water flow in a duct with a width $W(x)$ gradually varying along the direction of the flow with the spatially varying depth $H(x)$ and bottom profile $z_B = B(x)$ as shown in figure \ref{f00} (note that the water surface is not horizontal in the presence of spatially inhomogeneous flow). 
In a stationary flow, the current velocity $U(x)$ is related to the duct parameters by the flux conservation law:
 \be
 \Phi \equiv U(x)H(x)W(x) = \mbox{const}
 \label{Flux}
 \ee
and the Bernoulli equation:
 \be
 \frac{1}{2}\,U^2(x) + g\Bl[H(x) + B(x)\Br] = {\rm const},
 \label{Bern}
 \ee
where $g$ is the acceleration due to gravity. 
It is easy to see that by appropriately choosing profiles $B(x)$ and $W(x)$ one can provide the desired (and independent) variation along the canal of the flow velocity $U(x)$ and the speed of long waves $c(x) = \sqrt{gH(x)}$; this velocities are assumed to be positive everywhere.
%                   Fig. 0

 In the shallow-water theory, the linearised Euler equation is:
 \be
 \dfrac{\pl\tilde{u}}{\pl t}+\dfrac{\pl(U\tilde{u})}{\pl x}=-g\dfrac{\pl\eta}{\pl x},
 \label{Euler}
 \ee
 where $\tilde{u}(x,t)$ is the perturbation of the longitudinal component of fluid velocity, and $\eta(x,t)$ is the deviation of a free surface from the equilibrium state. Then the mass balance equation is:
 \be
 \dfrac{\pl S}{\pl t} + \dfrac{\pl}{\pl x}\Bl[S(U+\tilde{u})\Br] = 0,
 \label{Mass}
 \ee
 where $S(x,t)=[H(x)+\eta(x,t)]W(x)$ is a part of the duct cross-section occupied by water. Linearising this equation with respect to small perturbations $\eta$ and $\tilde{u}$ and taking into account equation (\ref{Flux}) with $\Phi \ne 0$, we obtain:
 \be 
 \dfrac{\pl\eta}{\pl t} + \dfrac{1}{W}\,\dfrac{\pl}{\pl x}\Bl[W(U\,\eta + H\,\tilde{u})\Br] \equiv \,
 \dfrac{\pl\eta}{\pl t} + HU\dfrac{\pl}{\pl x}\left(\dfrac{\eta}{H} + \dfrac{\tilde{u}}{U}\right) = 0.
 \label{eta}
 \ee

Introducing the velocity potential $\vp$ such that $\tilde{u} = \pl\vp/\pl x$, we integrate the linearised Euler equation (\ref{Euler}) and present $\eta$ in terms of $\vp$; then, a combination of equations (\ref{Euler}) and (\ref{eta}) gives the equation describing propagation of long wave:
 \[
 \left(\dfrac{\pl}{\pl t}+U\dfrac{\pl}{\pl x}-2U\dfrac{c'}{c}\right) \left(\dfrac{\pl\vp}{\pl t}+U\dfrac{\pl\vp}{\pl x} \right) = \dfrac{g}{W}\,
 \dfrac{\pl}{\pl x}\left(W H\,\dfrac{\pl\vp}{\pl x}\right),
 \]
where prime stands for the derivative with respect to $x$. This equation remains valid in the limiting case when $U(x) \equiv 0$. Using equation (\ref{Flux}) with $\Phi \ne 0$, we rewrite it in the form:
 \be
 \left(\dfrac{\pl}{\pl t}+U\dfrac{\pl}{\pl x}-2U\dfrac{c'}{c}\right)
 \left(\dfrac{\pl\vp}{\pl t}+U\dfrac{\pl\vp}{\pl x}\right)=
 c^2U\dfrac{\pl}{\pl x}\left(\dfrac{1}{U}\dfrac{\pl\vp}{\pl x}\right).
 \label{WEq0}
 \ee
This equation represents a basis for our analysis presented below but it does not admit the transition to the limiting case $U(x) \equiv 0$.

For what follows, it is convenient to present the velocity potential in the form: $\vp(x,t) = a(x)\psi(x,t)$, where the ``amplitude factor'' $a(x)$ will be defined later. Such a well-known trick allows one to reduce a linear equation with variable coefficients to one of the reference equations (see, e.g., \citep{Grimshaw-10, Pelinovsky-17} and references therein). After substitution of expression for $\varphi$ in equation (\ref{WEq0}), we arrive at the following equation for $\psi(x,t)$:
 \be
 \ba{l}
 \dfrac{\pl^2\psi}{\pl t^2} + (U^2-c^2)\,\dfrac{\pl^2\psi}{\pl x^2} +
 2U\,\dfrac{\pl^2\psi}{\pl t\pl x} +
 2U\,\left(\dfrac{a'}{a} - \dfrac{c'}{c}\right)\dfrac{\pl\psi}{\pl t}
   \\ \\ \phantom{wwa}
 +\,\left[2(U^2-c^2)\,\dfrac{a'}{a} + (U^2+c^2)\,\dfrac{U'}{U} -
 2U^2\dfrac{c'}{c}\right]\dfrac{\pl\psi}{\pl x} + Z(x)\psi = 0,
 \ea
 \label{WEq1}
 \ee
 where $Z(x)$ is defined by the equation:
 \be
 a(x)\,Z(x) = (U^2-c^2)a'' + a'\Bl[(U^2+c^2)(\ln U)' -
 2U^2(\ln c)'\Br].
 \label{Term}
 \ee
 
 Let us consider such a class of fluid flows for which $Z(x) \equiv 0$. Then, from (\ref{Term}) we obtain the exact differential equation for $a(x)$:
 \[
 \dd\Bl(\ln a'\Br) = \dd\left(\ln\,\dfrac{c^2U}{c^2-U^2}\right).
 \]
The first integral of this equation is:
 \be
 \dfrac{\dd a}{\dd x} = D\dfrac{\,c^2U}{c^2-U^2},
 \label{Eq-a}
 \ee
where $D$ is a constant of integration.

 Let us consider now a model equation:
 \be
 \left(\dfrac{\pl}{\pl t} + s_1(x)\,\dfrac{\pl}{\pl x} +F(x)\right)
 \left(\dfrac{\pl}{\pl t} + s_2(x)\,\dfrac{\pl}{\pl x}\right)f(x,t) = 0,
 \label{ModEq}
 \ee
 where $s_1(x)$, $s_2(x)$, and $F(x)$ are yet undefined functions. One of the solutions to this equation describes a travelling wave:
 \[
 f(x,t) = f_1\left(t - \int\dfrac{\dd x}{s_2(x)}\right),
 \]
 where $f_1(z)$ is an arbitrary function. Expansion of equation (\ref{ModEq}) leads to:
 $$
  \dfrac{\pl^2 f}{\pl t^2} + s_1(x)s_2(x)\,\dfrac{\pl^2 f}{\pl x^2} +
 \Bl[s_1(x) + s_2(x)\Br]\dfrac{\pl^2 f}{\pl t\pl x}
$$
 \be
{} + F(x)\,\dfrac{\pl f}{\pl t} + \Bl[s_1(x)s'_2(x) + F(x)s_2(x)\Br]
 \dfrac{\pl f}{\pl x} = 0.
 \label{ModEq1}
 \ee
 
 Let us compare this equation with (\ref{WEq1}) provided that $Z(x) \equiv 0$ and equation (\ref{Eq-a}) is hold. These equations are identical if we define functions $s_1(x)$, $s_2(x)$, and $F(x)$ such that:
$$
s_1(x)s_2(x) = U^2(x) - c^2(x), \quad s_1(x) + s_2(x) = 2U(x),
$$
\be
  F(x) = 2U(x)\,\left(\dfrac{a'(x)}{a(x)} - \dfrac{c'(x)}{c(x)}\right)
 \label{R1}
 \ee
 and
 $$
 s_1(x)s'_2(x) + F(x)s_2(x) 
 $$
 \be
 {} = 2\Bl[U^2(x) - c^2(x)\Br]\dfrac{a'(x)}{a(x)} +
 \Bl[U^2(x) + c^2(x)\Br]\dfrac{U'(x)}{U(x)} - 2U^2(x)\,\dfrac{c'(x)}{c(x)}\,.
 \label{R2}
 \ee

The first two equations (\ref{R1}) are hold if either $s_1 = U - c$, $s_2 = U + c$, or $s_1 = U + c$, $s_2 = U - c$. It is easy to see that in both these cases equation (\ref{R2}) yields (up to an unimportant numerical factor)
 \be
 a(x) = \Bl[c(x)U(x)\Br]^{1/2}.
 \label{a}
 \ee
Using this relation, we can rewrite equation (\ref{Eq-a}) in terms of functions $c(x)$ and $U(x)$ (positive values of the roots $c^{1/2}(x)$ and $U^{1/2}(x)$ are presumed hereafter):
 \be
\dfrac{\dd (c\,U)}{\dd x} \equiv c(x)\,\dfrac{\dd U}{\dd x} + U(x)\,\dfrac{\dd c}{\dd x} = 2D\,\dfrac{c^{5/2}(x)\,U^{3/2}(x)}{c^2(x) - U^2(x)}.
 \label{Eq-aa}
 \ee 

Thus, when the relationships (\ref{R1}), (\ref{a}) and (\ref{Eq-a}) are hold, equation (\ref{WEq0}) can be presented in one of the following forms:
 \begin{eqnarray}
{} &\phantom{\equiv}& \left[\dfrac{\pl}{\pl t} + \Bl(U - c\Br)\,\dfrac{\pl}{\pl x} +
 U\left(\dfrac{U'}{U} - \dfrac{c'}{c}\right)\right]
 \left[\dfrac{\pl}{\pl t}+ \Bl(U+c\Br)\dfrac{\pl}{\pl x}\right]\psi \nonumber\\
{} &\equiv& \left[\dfrac{\pl}{\pl t} + \Bl(U + c\Br)\,\dfrac{\pl}{\pl x} +
 U\left(\dfrac{U'}{U} - \dfrac{c'}{c}\right)\right]
 \left[\dfrac{\pl}{\pl t} + \Bl(U-c\Br)\dfrac{\pl}{\pl x}\right]\psi = 0.
  \label{WEq2}
 \end{eqnarray}
 The general solution of this equation can be presented as a superposition of two waves of an arbitrary form travelling with the different velocities: 
  \be
 \psi(x,t) = \psi_1\left(t-\int\dfrac{\dd x}{U(x)+c(x)}\right) +
 \psi_2\left(t-\int\dfrac{\dd x}{U(x)-c(x)}\right).
 \label{psi}
 \ee
 The independent propagation of each of these waves in the inhomogeneous fluid is provided by the single condition (\ref{Eq-a}) which relates the fluid and wave speeds, $U(x)$ and $c(x)$. 
 Therefore, equation (\ref{Eq-a}) when it holds provides the reflectionless (RL) wave propagation. 
 
 When the functions $a(x)$ and $\psi(x, t)$ are found, then we get the velocity potential $\varphi$ which allows as to find the velocity $u$ and water surface elevation $\eta$:
$$  
\varphi(x,t) = a(x)\psi(x,t), \qquad
u = \frac{\partial\varphi}{\partial x}, \qquad \eta = -\frac{1}{g}\left(\frac{\partial\varphi}{\partial t} + U\frac{\partial\varphi}{\partial x}\right). 
$$
 
 A wide class of various RL flows can be obtained even in the simplest case when $D = 0$. In such a case, (\ref{Eq-a}) reduces to the condition $a(x) = \mbox{const}$ or, equivalently (see, for example, \citep{ChSt21}):
 \be
 c(x)U(x) = \mbox{const}.
 \label{M1}
 \ee
%\sout{which was obtained earlier (see, for example, \citep{ChSt21}).} 
The profile of one of these speeds can be specified as an arbitrary continuous positively-defined function, and the profile of another speed is uniquely calculated then. It is important to note that, firstly, such a flow is global, i.e. is defined on the entire $x$-axis, and, secondly, neither the presence nor absence of critical points in the flow, at which the velocities $U(x)$ and $c(x)$ become equal, does not affect in any way such properties as smoothness and boundedness of the velocities.
 \section{Reflectionless flows in the case $D \ne 0$. A general analysis}
\label{Sect03}
If $D \ne 0$ being positive or negative constant, singularities appear in the right-hand side of equation (\ref{Eq-aa}) when either $c = U$, or $U = 0$, or $c = 0$, and also when $U(x)$ or $c(x)$ grow with no limit when $x$ approaches some point $x_*$. If at least one of these singularities occurs at a finite point $x = x_*$, then several questions arise: (1) about the continuation of the flow model through this point, (2) about the contribution of the singular point to the wave reflection and/or absorption processes, and, more broadly, (3) about the existence of global RL flows. Below we investigate these issues.

We note first that equations (\ref{Eq-a}) and (\ref{Eq-aa}) are invariant with respect to the simultaneous replacement $x \to -x$ and $D \to -D$. For this reason, we introduce a variable $\xi = Dx$ which remains the same under such a transformation and rewrite equations (\ref{Eq-a}) and (\ref{Eq-aa}) as:
\be
 \dfrac{\dd a(\xi)}{\dd\xi} = \dfrac{c^2(\xi)U(\xi)}{c^2(\xi)-U^2(\xi)}
 \quad {\rm and} \quad
 c(\xi)\dfrac{\dd U(\xi)}{\dd\xi} + U(\xi)\dfrac{\dd c(\xi)}{\dd\xi} =
 \dfrac{2\,c^{5/2}(\xi)U^{3/2}(\xi)}{c^2(\xi)-U^2(\xi)}\,.
 \label{Eq-a1}
 \ee
In addition, equations (\ref{Eq-a}), (\ref{Eq-aa}), and (\ref{Eq-a1}) are invariant with respect to the scaling transformation $U(x)\ \to\ U(x)/c_s$,\ \ $c(x)\ \to\ c(x)/c_s$, where $c_s = \mbox{const}$. In what follows, we will choose the appropriate scale $ c_s $. Note also that these equations possess the translational symmetry which means that if $c(\xi)$ and $U(\xi)$ satisfy equation (\ref {Eq-a1}), then $c(\xi + b)$ and $U(\xi + b)$, where $b$ is an arbitrary constant, satisfy this equation too.

Let us introduce the dimensionless functions: $u(\xi) = w^{-1}(\xi) = \Bl[U(\xi)/c(\xi)\Br]^{1/2}$, noting that $c(\xi)u(\xi)=\Bl[c(\xi)U(\xi)\Br]^{1/2}\equiv a(\xi)$. Then, let us re-write equation (\ref{Eq-a}) in the forms:
 \be
 \dfrac{\dd u}{\dd\xi} = \dfrac{u^2}{1-u^4} - \dfrac{u}{c(\xi)}\,
 \dfrac{\dd c(\xi)}{\dd\xi}
 \label{u1}
 \ee
 and
 \be
 \dfrac{\dd}{\dd\xi}\,\ln a(\xi) = \dfrac{u(\xi)}{1-u^4(\xi)} =
 \dfrac{w^3(\xi)}{w^4(\xi)-1}\,.
 \label{auw}
 \ee
The right-hand side of equation (\ref{auw}) is positive when $u < 1$ ($w > 1$) and
 is negative when $u > 1$ ($w < 1$), therefore, $a(\xi)$ increases monotonically in the subcritical flow and decreases monotonically in the supercritical flow. If
 we assume that $c(\xi)$ is a bounded function at any finite $\xi$, i.e.
 $0 < c(\xi) < \infty$, then the singularities at $U=0$ and $U=\infty$ can be reached only asymptotically when $\xi \to -\infty$. Similarly, under the assumption of the boundness of $U(\xi)$, singularities at $c = 0$ and $c = \infty$ can  be reached only asymptotically when $\xi \to +\infty$. In all these cases, the right-hand side of equation (\ref{auw}) vanishes, hence the derivative in the left-hand side vanishes too. As the result of this, when $|\xi| \to \infty$, function $a(\xi)$ either asymptotically approaches some constant (which may be zero, in particular) or grows slower than an exponential function, e.g., in a power-type manner, $a(\xi)\sim |\xi|^\al$, where $\al>0$. In contrast to this,
 a singularity at $U=c$\ \ ($u=w=1$) can be achieved (as will be shown below) at a finite point $\xi$.

 To conclude this section, we consider the behaviour of the functions $U(\xi)$ and $c(\xi)$
 in the neighborhood of a critical point where $U = c$. Assume that $\xi=0$ is a critical point, and the normalisation constant $c_s$ is chosen such that $U(0) = c(0) = 1$. Let us set
 \[
 c(\xi) = 1 + s(\xi), \qquad U(\xi) = 1 + v(\xi),
 \]
 where $s(\xi)\to 0$ and $v(\xi)\to 0$ when $\xi\to 0$. After substitution in the second equation of (\ref{Eq-a1}), we obtain:
 \[
 \Bl(s - v\Br)\Bl(s' + v'\Br) = 1 + 2s + v + O(s^2 + v^2),
 \]
where prime stands for the derivative with respect to $\xi$. From here, we see that $(s-v)$ and $(s+v)$ are power-type functions of $\xi$:
 \be
 (s - v) \sim \xi^\al, \qquad (s + v) \sim \xi^\beta,
 \label{albet}
 \ee
where $\al + \beta = 1$, $\al > 0$, $\beta > 0$.

Since $\xi=0$ is a branching point and $s(\xi)$ and $v(\xi)$ 
 are real functions, the cases $\xi>0$ and $\xi<0$ should be considered
 separately. \\
 
 {\bf Case A.} If $\al=\beta=1/2$, then we readily find for $\xi\ge 0$:
 \be
 c(\xi) = 1+s_+\xi^{1/2}+O(\xi), \qquad U(\xi) = 1+v_+\xi^{1/2}+O(\xi),
 \label{as+}
 \ee
 where $s_+^2 = v_+^2 + 2$.
 
For $\xi\le 0$, we find: 
 \be
 c(\xi) = 1+s_-(-\xi)^{1/2}+O(\xi), \qquad U(\xi) = 1+v_-(-\xi)^{1/2}+O(\xi),
 \label{as-}
 \ee
 where $s_-^2 = v_-^2 - 2$. Here $s_\pm$ and $v_\pm$ are constants which can have any sign, depending on the branch of the solution (subcritical or supercritical). Moreover, it is not necessary that the both
 velocities have singularities at $\xi=0$. In particular, in the domain where $\xi \ge 0$, the flow velocity can be regular ($v_+=0$), while the wave velocity can be regular ($s_- =0$) in the domain where $\xi \le 0$ .

 If $\al\ne\beta$ then, we can present $s(x)$ and $u(x)$ as:
  \[
 s(\xi) = s_{a\pm} |\xi|^\al + s_{b\pm} |\xi|^\beta + \dots, \qquad
 v(\xi) = v_{a\pm} |\xi|^\al + v_{b\pm} |\xi|^\beta + \dots,
 \]
 where the indices plus (minus) pertain to the regions $\xi\ge 0$ ($\xi\le 0$). \\
 
  {\bf Case B.} Consider the case when $0<\al<\beta$, then from the condition $(s+v)\sim |\xi|^\beta$, we obtain:
 \be
 s_{a\pm}+v_{a\pm} = 0, \qquad 2\beta s_{a\pm}(s_{b\pm}+v_{b\pm}) = \pm 1.
 \label{ab}
 \ee \\

 {\bf Case C.} In another case, $\al>\beta>0$, from the condition $(s-v)\sim |\xi|^\al$, we obtain:
 \be
 s_{b\pm} = v_{b\pm}, \qquad 2\beta s_{b\pm}(s_{a\pm} - v_{a\pm}) = \pm 1.
 \label{ba}
 \ee
 Note that if $\al \ne \beta$, then both velocities inevitably have a singularity in the point $\xi=0$.

 \section{Examples of reflectionless flows}
\label{Sect04}
\subsection{Currents with a constant wave speed}
\label{Sect041}
\begin{figure}
\vspace{-3.5cm}
\centerline{\includegraphics[width=1.0\textwidth]{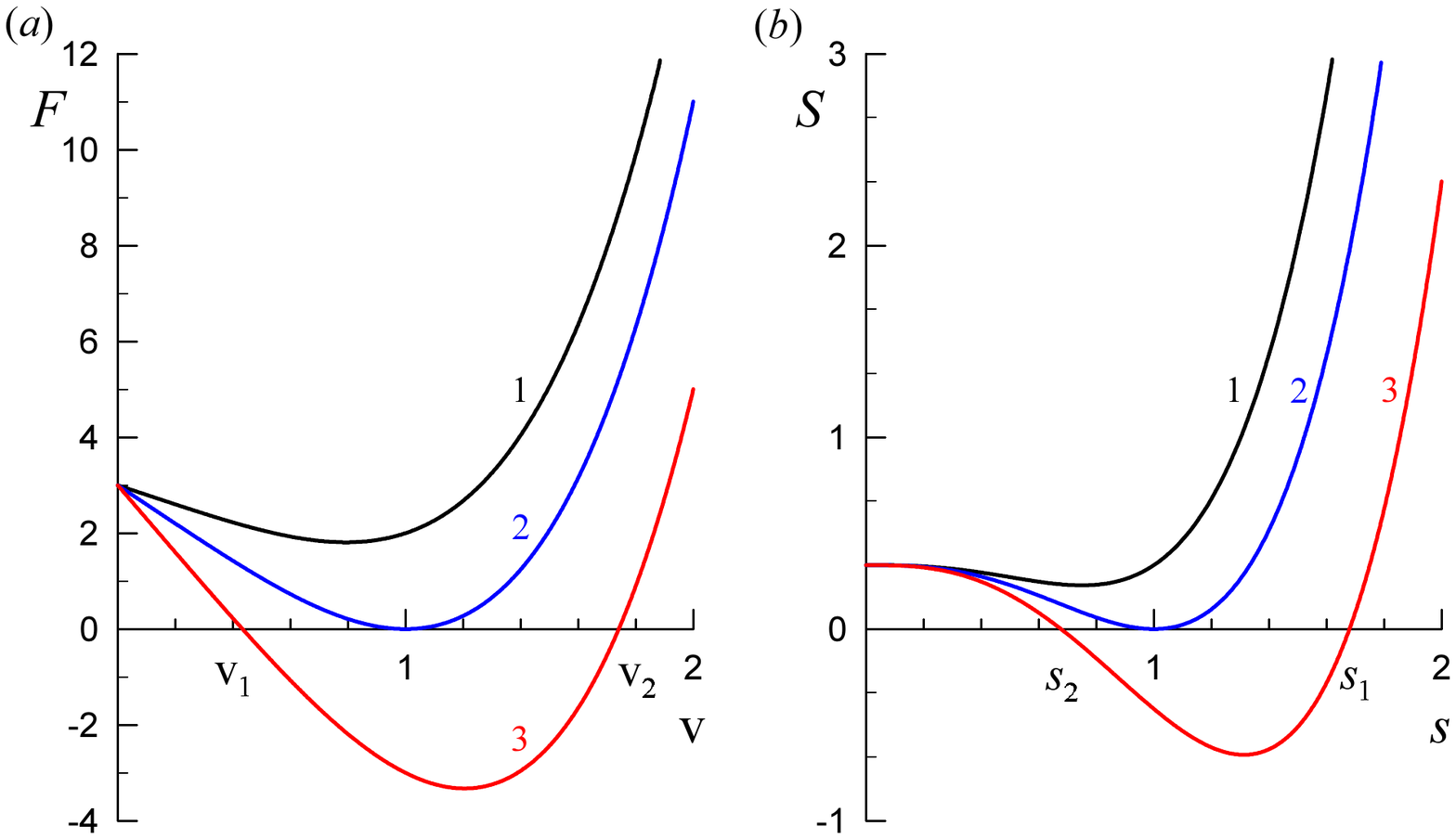}}%
\vspace{-7.5cm}
\caption{(Color online.) (a) -- Roots of equation (\ref{v}) for $\xi_0-\xi =2/3$ (line 1); $\xi_0-\xi = 4/3$ (line 2); $\xi_0-\xi = 7/3$ (line 3). (b) Roots of equation (\ref{s}) for $\xi-\xi_0 = 1$ (line 1); $\xi-\xi_0 = 4/3$ (line 2); $\xi-\xi_0 = 1.75$ (line 3).}
\label{f01}
\end{figure}

\begin{figure}
\vspace{-1.0cm}
\centerline{\includegraphics[width=0.8\textwidth]{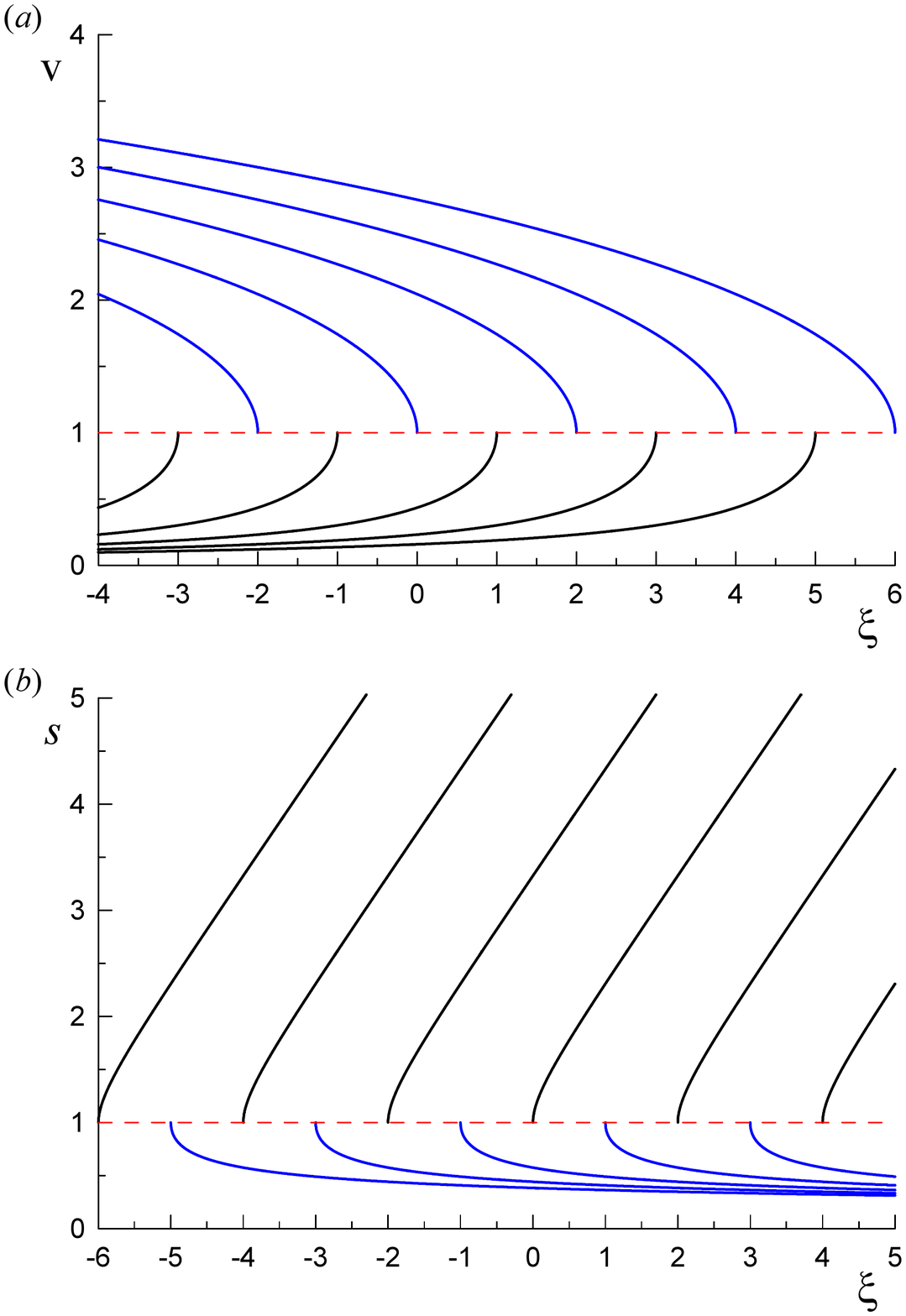}}%
\vspace{-0.5cm}
\caption{(a) Solutions of equation (\ref{v}) for the different values of $\xi_0$
 in the subcritical (lover part of the frame) and supercritical (upper part of the frame) regions. (b) The same for the solutions of equation (\ref{s}), but graphs for subcritical flows ($s>1$) are in the upper part, whereas graphs for supercritical flows ($s<1$) are in the lower part of the frame.}
\label{f02}
\end{figure}

As a first example, consider a current with a constant wave speed $c(\xi)\equiv 1$. From equation (\ref{Eq-a1}) we find:
 \[
 \dfrac{U'(\xi)}{U^{3/2}(\xi)}\Bl[1-U^2(\xi)\Br] = 2 \ \ \Longrightarrow \ \
 U^2(\xi) + 3(\xi-\xi_0)U^{1/2}(\xi) + 3 = 0.
 \]
Using the substitution $v(\xi) = U^{1/2}(\xi)$, we arrive at the algebraic equation for $v$:
 \be
 F(v) \equiv v^4 + 3(\xi-\xi_0)v + 3 = 0.
 \label{v}
 \ee
This equation has two positive real roots for $\xi < \xi_* = \xi_0-4/3$, one of them $v_1(\xi) < 1$ and another one $v_2(\xi) > 1$ (see figure \ref{f01}(a)). The roots merge into one double root $v=1$ (corresponding to the critical point where $U = c = 1$) when $\xi = \xi_*$. If $\xi > \xi_*$, then the polynomial (\ref{v}) does not have real roots. Without loss of generality, we can set $\xi_0 = 4/3$ ($\xi_* = 0$) bearing in  mind that all the other solutions can be obtained by a shift in $\xi_0$. Then for $0 \le -\xi \ll 1$ we have:
\be
 v_{1,2}(\xi) = 1 \mp \left(-\frac{\xi}{2}\right)^{1/2} -\frac{\xi}{12} +
 \dots, \quad
 U_{1,2}(\xi) = 1 \mp \Bl(-2\xi\Br)^{1/2} - \frac{2}{3}\,\xi \dots ,
 \label{xi-0}
 \ee
 and for $\xi \to -\infty$ the asymptotic solutions originated for each root are (we are recalling that $a(\xi)=c(\xi)\,v(\xi)$):
 \be
 \ba{ll}
 a_1(\xi) = v_1(\xi) \approx \Bl(4/3 - \xi\Br)^{-1}, \ \ & \ \
 a_2(\xi) = v_2(\xi) \approx \Bl(4 - 3\xi\Br)^{1/3}, \\
 U_1(\xi) \approx \Bl(4/3 - \xi\Br)^{-2}, \ \ & \ \
 U_2(\xi) \approx \Bl(4 - 3\xi\Br)^{2/3}.
 \ea
 \label{xi-inf}
 \ee
%                   Fig. 1
The subcritical solution ($U_1$) and supercritical solution ($U_2$) are defined only for the negative $\xi\le 0$. Profiles $v(\xi)$ are shown in figure \ref{f02}(a) for different values of $\xi_0$. With the known dependences of $U_{1,2}(\xi)$ and fixed water depth $H_0$, we can calculate the dependences of the duct width $W(\xi)$ and bottom profile $z_B = B(\xi)$. Using equations
(\ref{Flux}) and (\ref{Bern}), we obtain:
\be
\label{FloPar-1}
W(\xi) \sim U^{-1}(\xi), \qquad B(\xi) = B(0) - \frac{1}{2g}U^2(\xi).
\ee
%                   Fig. 2
\subsection{Currents with a constant flow speed}
\label{Sect042}
 A very similar features has a flow with a constant current speed, $U(\xi)=1$. In this case equation (\ref{Eq-a1}) reduces to:
 \[
 \dfrac{c'(\xi)}{c^{5/2}(\xi)}\Bl[c^2(\xi)-1\Br] = 2 \ \ \Longrightarrow \ \
 c^2(\xi) - (\xi-\xi_0)c^{3/2}(\xi) + \frac{1}{3} = 0,
 \]
 or equivalently to:
 \be
 S(s) \equiv s^4 - (\xi-\xi_0)s^3 + \frac{1}{3} = 0, 
 \label{s}
 \ee
where $s(\xi) = c^{1/2}(\xi)$. There are two positive real roots for $\xi>\xi_* = \xi_0+4/3$, $s_1(\xi)>1$ and $s_2(\xi)<1$ corresponding to the subcritical and supercritical flows, respectively. These roots merge at $s=1$ when $\xi = \xi_*$. The polynomial (\ref{s}) does not have real roots if $\xi < \xi_*$ (see figure \ref{f01}(b). Setting $\xi_0 = -4/3$
 ($\xi_* = 0$), we find for $0\le \xi\ll 1$:
 \be
 s_{1,2}(\xi) = 1 \pm \left(\frac{\xi}{2}\right)^{1/2} +
 \frac{5}{12}\,\xi + \dots,
 \quad c_{1,2}(\xi) = 1 \pm \Bl(2\xi\Br)^{1/2} +
 \frac{4}{3}\,\xi + \dots
 \label{xi-0c}
 \ee
Then, for $\xi\to +\infty$, we have (here $a(\xi)=s(\xi)\,U(\xi)$):
 \be
 \ba{ll}
 a_1(\xi) = s_1(\xi) \approx \xi + 4/3, \ & \
 a_2(\xi) = s_2(\xi) \approx \Bl(3\xi + 4\Br)^{-1/3}, \\
 c_1(\xi) \approx \Bl(\xi + 4/3\Br)^2, \ & \
 c_2(\xi) \approx \Bl(3\xi + 4\Br)^{-2/3}.
 \ea
 \label{xi-infc}
 \ee
  Note that, in contrast to the previous example, the solutions are defined only on the positive semi-axis $\xi \ge 0$, because at $\xi=0$ function $U(\xi)$ is regular.
 The profiles of $s(\xi)$ are shown in figure \ref{f02}(b) for different values of $\xi_0$. From the known dependences of $c_{1,2}(\xi) = \sqrt{gH(\xi)}$, we immediately derive the dependence of water depth $H_{1,2}(\xi) = c^2_{1,2}(\xi)/g$. Then, from equation (\ref{Flux}) with the fixed current flow $U_0$, we can calculate the dependences of the duct width $W(\xi)$ and bottom profile $z_B = B(\xi)$:
\be
\label{FloPar-2}
W(\xi) \sim H^{-1}(\xi)  \sim c^{-2}(\xi), \qquad B(\xi) = - H(\xi) + \mbox{const.}
\ee

In the next examples, we consider flows in which both velocities are unlimited in general.
\subsection{Currents with functionally related velocity profiles}
\label{Sect043}
 Consider now such class of fluid flows in which $U(\xi)$ and $c(\xi)$ are related by the power type equation:
 \be
 U(\xi) = Vc^{p}(\xi), \quad V = {\rm const} > 0,\ \ p = {\rm const}.
 \label{lb}
 \ee
 We assume that $p \ne \pm 1$. When $p=-1$, we arrive at the equation $c(x)U(x) = \mbox{const.}$ which corresponds to the case $D = 0$ (see the end of Section \ref{Sect02}). When $p=1$, we obtain a flow with the constant Froude number, $\mbox{Fr} \equiv U(\xi)/c(\xi) = V$, but with the varying $c(\xi)$ and $U(\xi)$:
 \be
 c(\xi) = C\re^{\mu \xi}, \quad U(\xi) = Vc(\xi),\ \
 \mu = \dfrac{V^{1/2}}{1-V^2}\,.
 \label{M2}
 \ee
 As will be shown below, these are precisely such solutions that serve as the separatrices, separating solutions with the fundamentally different behaviours.

 Substituting $U(\xi)$ from  (\ref{lb}) into equation (\ref{Eq-a1}) and integrate it, we obtain:
 \[
 \dfrac{2}{1-p}\,c^{(1-p)/2} + \dfrac{2V^2}{3(1-p)}\,c^{3(p-1)/2}
 = \dfrac{2V^{1/2}}{1+p}\Bl(\xi-\xi_0\Br).
 \]
Introduction of new variable $u>0$ such that $u^2 \equiv U/c = Vc^{p-1}$, yields:
 \be
 u^4 + 3D_p(\xi-\xi_0)\,u + 3 = 0, \qquad D_p=\dfrac{p-1}{p+1}\,.
 \label{u}
 \ee
 In terms of the variable $\zeta=D_p\xi$, this equation takes the same form as equation (\ref{v}), therefore, its solutions are the same, $u_{1,2}(\zeta) \equiv v_{1,2}(\zeta)$ (see figure \ref{f02}(a)). All other variables $c$, $U$ and $a$ expressed in terms of $u$ are:
 \be
 c = c_*\,u^{\frac{2}{p-1}}, \qquad
 U = u^2\,c = c_*\,u^{\frac{2p}{p-1}}, \qquad
 a = u\,c = c_*\,u^{\frac{p+1}{p-1}},
 \label{cUu}
 \ee
 where $c_*= V^{\frac{1}{1-p}}$ is the value of velocities at the critical point $\xi = \xi_*$,\ \ $U(\xi_*) = c(\xi_*) = c_*$.
 As one can see, the behaviour of these  functions depend on the parameter $p$. Indeed, if we set $\zeta_0\equiv D_p\xi_0=4/3$, then we find with the help of equations (\ref{xi-inf}) and (\ref{cUu}) that when $\zeta\to-\infty$, the asymptotic behaviour of the functions of interest are:
 \be
 \ba{lll}
 c_1 \approx c_*\Bl(-\zeta\Br)^{\frac{2}{1-p}}, \ & \
 U_1 \approx c_*\Bl(-
 \zeta\Br)^{\frac{2p}{1-p}}, \ & \
 a_1 \approx c_*\Bl(-
 \zeta\Br)^{\frac{1+p}{1-p}}, \\ \\
 c_2 \approx c_*\Bl(- 3\zeta\Br)^{\frac{2}
 {3(p-1)}}, \ & \
 U_2 \approx c_*\Bl(- 3\zeta\Br)^{\frac{2p}
 {3(p-1)}}, \ & \
 a_2 \approx c_*\Bl(- 3\zeta\Br)^{\frac{p+1}
 {3(p-1)}}.
 \ea
 \label{cU1}
 \ee
From this expressions we derive:
 \be
 \ba{lll}
 c_1,\ U_1,\ a_1\ \to 0,\ & \ c_2,\ U_2,\ a_2\ \to \infty,\ \ & \ \
 {\rm for}\ p > 1,
  \\
 c_1,\ U_1,\ a_1\ \to \infty,\ & \ c_2,\ U_2,\ a_2\ \to 0,\ \ & \ \
 {\rm for}\ 0 < p < 1,
   \\
 c_1,\ a_1 \to \infty,\ U_1\ \to 0,\ & \ c_2,\ a_2 \to 0,\ U_2\ \to \infty,
 \ \ & \ \ {\rm for}\ -1 < p < 0,
   \\
 c_1 \to \infty,\ U_1,\ a_1\ \to 0,\ & \ c_2 \to 0,\ U_2,\ a_2\ \to \infty,
 \ \ & \ \ {\rm for}\ p < -1.
 \ea
 \label{cU2}
 \ee
 In the neighbourhood of the critical point ($\zeta \to 0_-$) we have:
 \[
 c_{1,2}(\zeta) = c_*\left[1 \mp
 \dfrac{(-2\zeta)^{1/2}}{p-1} + \dots\right], \qquad
 U_{1,2}(\zeta) = c_*\left[1 \mp
 \dfrac{p}{p-1}\Bl(-2\zeta\Br)^{1/2} + \dots\right].
 \]
 
\begin{figure}
 \vspace{-2cm}
 \epsfxsize=140mm
 \centerline{\epsfbox{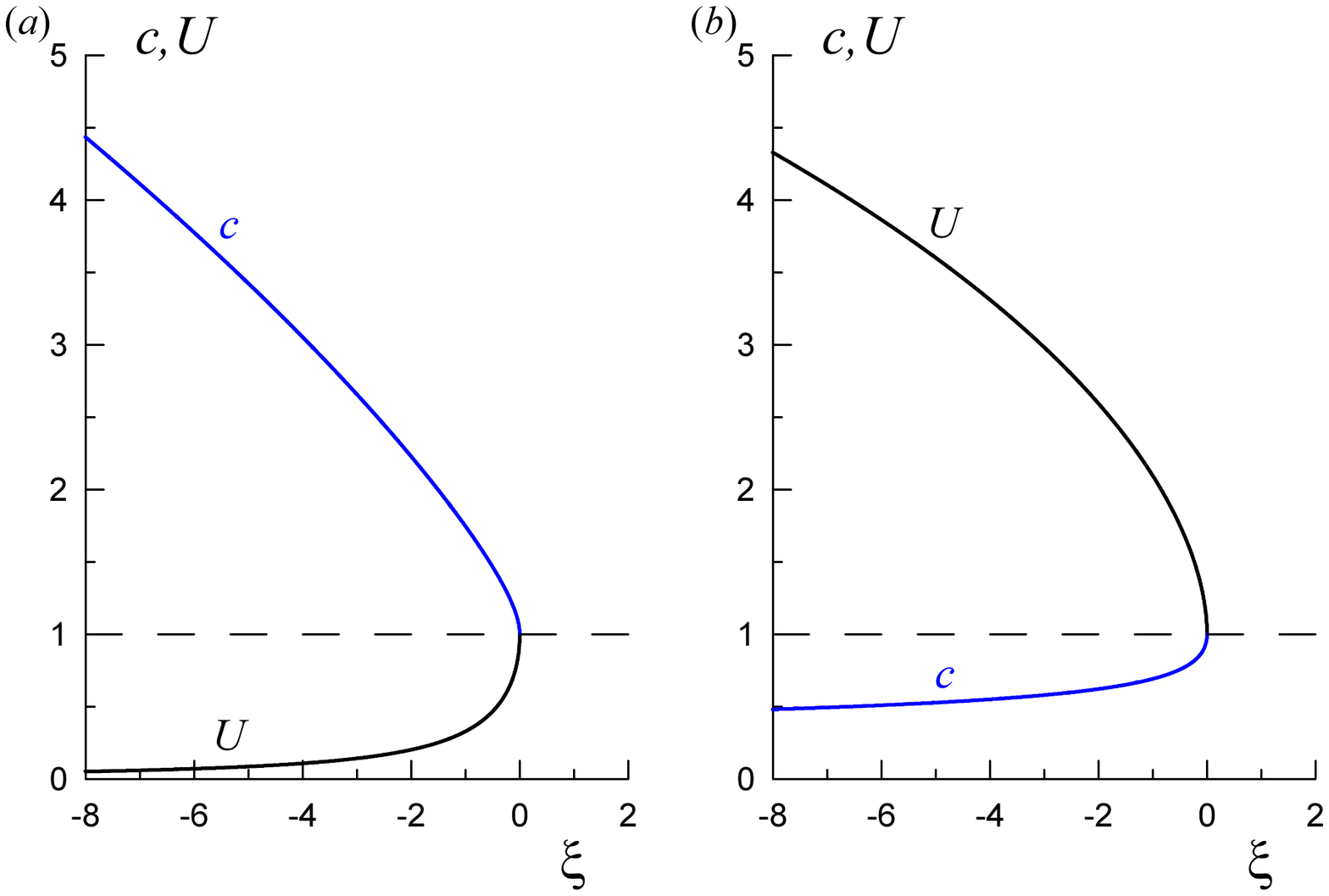}}
 \vspace{-8.5cm}
 \caption{(Color online.) The $c(\xi)$ and $U(\xi)$ profiles of the RL flows in a duct of a constant width: (a) subcritical flow and (b) supercritical flow.}
 \label{f08}
 \end{figure}

As an example, let us consider the flow in a duct of constant width $W$. In accordance with the flux conservation law (\ref{Flux}), such a flow corresponds to $p = -2$, and formulae (\ref{cUu}) yield:
\[
U(\xi) = c_*\,u^{4/3}(\xi), \qquad 
c(\xi) = c_*\,u^{-2/3}(\xi), \qquad H(\xi) = \dfrac{c^2_*}{g}\,u^{-4/3}(\xi).
\] 
The profiles of velocities for $\xi_* = 0$ and $c_*= 1$ are shown in figure~\ref{f08}.
%                   Fig. 8

It should be noted that in all models with $D \ne 0$ considered above, solutions are defined only on the semi-axis $\xi$, and the domains of their definition are limited because a singularity with $c = U$ occurs at a finite value of $\xi=\xi_*$. 
As the velocity $U(\xi_*) \ne 0$, it is evident that the flow goes over the point $\xi = \xi_*$ into the domain where $\xi > \xi_*$. 
One of the versions of such flow model is discussed below in Section \ref{Sect062}.
%From the physical point of view, as far as $U(\xi_*) \ne 0$, the flow must continue into the $\xi > \xi_*$ domain, and a version of such continuation is discussed below, in Section \ref{Sect062}.

%
\subsection{A special class of  flows with the exponential variation of wave speed $c(\xi)$}
\label{Sect044}
 The flows of this class differ significantly from those considered above. Let $c(\xi) = C\exp(\mu\xi)$. Then equation (\ref{u1}) takes the form:
 \be
 \dfrac{\dd u}{\dd\xi} = \dfrac{u^2}{1-u^4} - \mu u.
 \label{u2}
 \ee
 This equation, in contrast to the considered above, has a stationary point (the null-isocline) $u(\xi) = u_0 = \mbox{const.}$ in which its right-hand side vanishes. The value of $u_0$ is a positive root of the following equation (cf. with the last equation (\ref{M2}) which reduces to (\ref{u0}) if we set $V = u_0^2$):
 \be
 u_0^4 + \dfrac{u_0}{\mu} - 1 = 0.
 \label{u0}
 \ee
 It is easy to see that for $\mu > 0$, the value of $u_0 < 1$, i.e. corresponds to the 
  subcritical flows, whereas for $\mu < 0$, the value of $u_0 > 1$, which corresponds to
  supercritical flows.

  Substituting $u = u_0 + v$, where $|v| \ll u_0$, in equation (\ref{u2}), we obtain after linearisation with respect to $v$
 \[
 \dfrac{\dd v}{\dd\xi} = \left[\dfrac{2u_0}{1-u_0^4} + \dfrac{4u_0^5}
 {(1-u_0^4)^2} - \mu\right]v = \mu(1 + 4\mu u_0^3)v.
 \]
 Solutions to this equation,
 \be
 v(\xi) = v_0\exp(\gamma \xi), \quad \gamma = \mu(1 + 4\mu u_0^3), \quad v_0 = \mbox{const},
 \label{as-u0}
 \ee
 vanish when $\xi \to -\infty$. This is obvious when $\mu > 0$ and, therefore, $\gamma > 0$. When $\mu < 0$, we obtain with the help of equation (\ref{u0}) that $\gamma$ is still positive:
 \[
 \gamma = \mu\Bl[1 + 4(\mu/u_0 - 1)\Br] = 4\mu^2/u_0 - 3\mu >0.
 \]
 This means that the stationary point $u_0$ is unstable due to small perturbation $v(\xi)$ grows with $\xi$.

\begin{figure}
\vspace{-1.0cm}
\centerline{\includegraphics[width=0.8\textwidth]{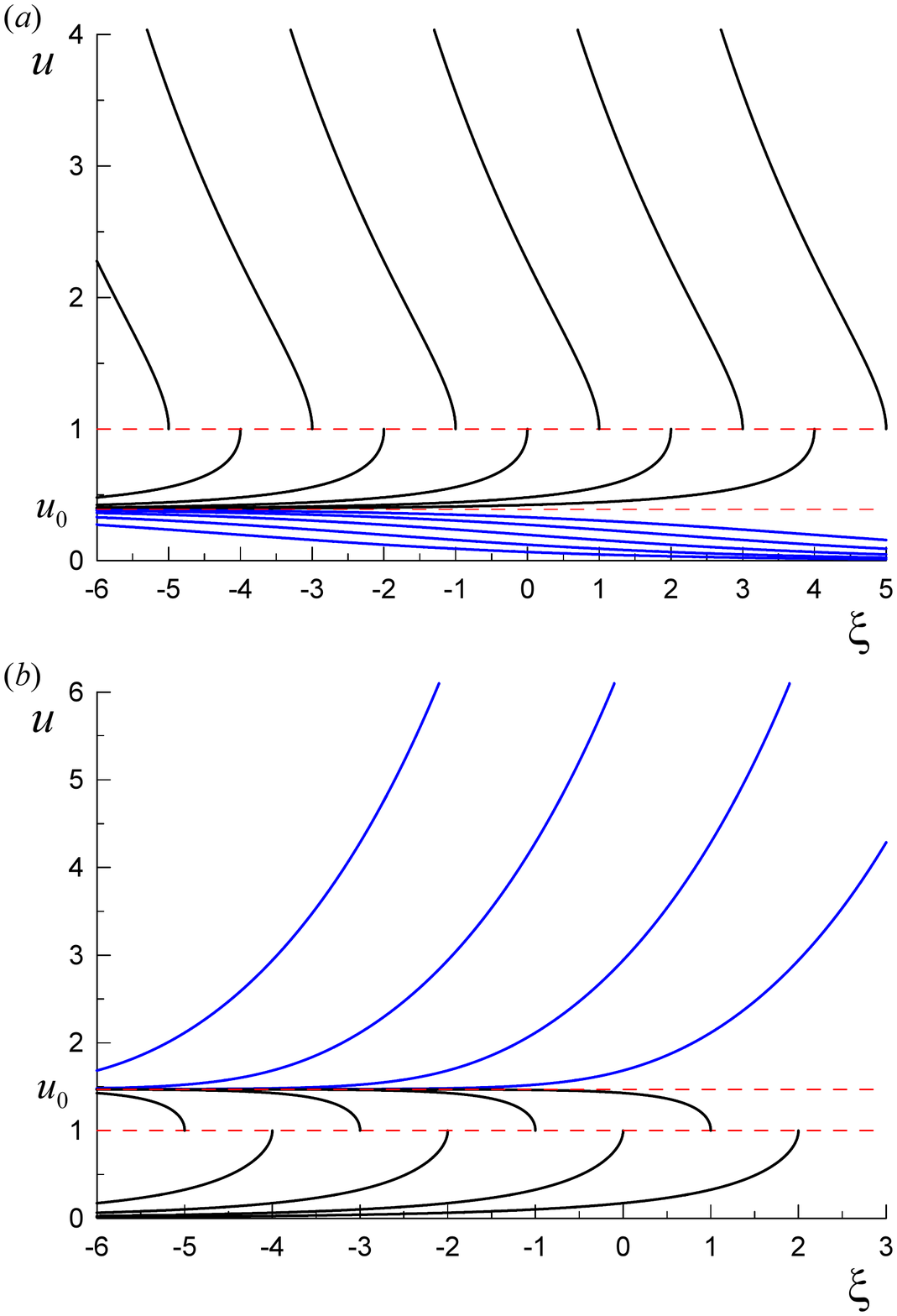}}%
\vspace{-0.5cm}
\caption{(Color online.) Solutions of equation (\ref{u2}). Frame (a) pertains to positive $\mu=0.4$ ($u_0=0.3907$); frame (b) pertains to negative $\mu=-0.4$ ($u_0=1.4705$). Global solutions are depicted by blue lines.}
\label{f03}
\end{figure}

As a result, the upper half-plane $(\xi, \,u)$ is split into three strips (separated from each other by the straight lines $u = u_0$ and $u = 1$), in which the behavior of solutions to equation (\ref{u2}) is significantly different (see figure \ref{f03}).
%                   Fig. 3

If $\mu > 0$, then $0 < u_0 < 1$, and solutions to equation (\ref{u2}) are such as shown in figure \ref{f03}(a). In the lower strip, $0 < u < u_0$, any solution $u(\xi)$ decreases monotonically approaching $u_0$ from below when $\xi \to -\infty$ and zero from the top when $\xi \to +\infty$, according to $u \approx u_{m+}\exp(-\mu \xi)$. Respectively, in the same limit we have:
 \be
 U(\xi) \equiv u^2(\xi)\,c(\xi) \approx Cu^2_{m+}e^{-\mu\xi}, \quad
 a(\xi) = \Bl[c(\xi)\,U(\xi)\Br]^{1/2} \to C\,u_{m+} = {\rm const}.
 \label{Ua}
 \ee
 
 In the middle strip, $u_0 < u \le 1$, any solution $u(\xi)$ grows monotonically from $u_0$ when $\xi \to -\infty$ to $u = 1$ at some $\xi_*$. Near this limiting coordinate $\xi_*$ the asymptotic solution is
 \be
 u(\xi) = 1 - \left(\frac{\xi_* - \xi}{2}\right)^{1/2} -
 \frac{1 - 8\mu}{12}\Bl(\xi_* - \xi\Br) + \dots
 \label{1-}
 \ee
 
 In the upper strip any solution tends to infinity as $u \approx u_{m-}\exp(-\mu\xi)$ when $\xi \to -\infty$ and decreases with $\xi$ approaching unity, when $\xi$ tends to some $\xi_+$ as per the formula:
  \be
 u(\xi) = 1 + \left(\frac{\xi_+ - \xi}{2}\right)^{1/2} -
 \frac{1 - 8\mu}{12}\Bl(\xi_+ - \xi\Br) + \dots
 \label{1+}
 \ee

If $\mu < 0$, then $u_0 > 1$. Solutions to equation (\ref{u2}) are shown in figure \ref{f03}(b). In the lower strip, $0 < u < 1$, all solutions $u(\xi)$ monotonically increase from $u\approx u_{M-}\exp(-\mu\xi)$ when $\xi \to -\infty$ attaining $u = 1$ at some finite $\xi$ according to the dependence similar to (\ref{1-}).  In the middle strip, $1 < u < u_0$, solutions $u(\xi)$ monotonically decrease from $u_0$ when $\xi \to -\infty$ to $u = 1$ achieving this value at some finite $\xi$ as per the dependence similar to (\ref{1-}). In the upper strip, $u > u_0$, solutions $u(\xi)$ monotonically increase from $u_0$ when $\xi \to -\infty$, to $u\approx u_{M+}\exp(-\mu\xi)$ when $\xi \to +\infty$; the dependences similar to (\ref{Ua}) are held in this limit.

It should be noted that in the lower strip for $\mu > 0$ and in the upper strip for $\mu < 0$ solutions describing flows are {\it global}, i.e., they are defined on the entire $\xi$-axis, whereas in other strips, solutions are defined only to the left of some finite point $\xi_*$ which is different for each particular realization.
 \section{Global solutions and conditions of their existence}
\label{Sect05}
As follows from the analysis presented above for the specific flow models, in the majority of cases solutions for the RL current profile are applicable only on the limited spatial interval, because the profiles of the duct depth, width, or current become either singular in certain points, or diverge at the infinity. Therefore, one of the important questions is whether it is possible to find such conditions when solutions for the profiles are globally defined on the entire $x$-axis. Such a problem was solved in the case when there is no current \citep{Pelinovsky-17}. We provide the solution for the case when the current is taken into account.

 Let us recall that in the case of the exponential variation of the wave speed $c(\xi)$ considered above, the null-isoclines are horizontal ($u_0 =\mbox{const.}$). Therefore, they serve as separatrices separating the global solutions from those that are defined on the bounded $x$-interval. Under a non-exponential variation of $c(\xi)$, the null-isocline $u = u_0(\xi)$, on which the right-hand side of equation (\ref{u1}) vanishes, is still described by equation (\ref{u0}) but now 
 \be
 \mu(\xi) \equiv \frac{\dd }{\dd \xi}\ln{c(\xi)} \ne \mbox {const}.
 \ee
 Therefore, it is not horizontal and is intersected on the $(\xi,\,u)$-plane by some solutions of equation (\ref{u1}). In what follows, we will assume that $c(\xi)$ either increases monotonically or decreases monotonically determining accordingly the sign of $\mu(\xi) \ne 0$. As follows from equation (\ref{u1}), when $\mu(\xi) > 0$, then only subcritical solutions ($u(\xi) < 1$) can be global, whereas when $\mu(\xi) < 0$, only supercritical solutions ($u(\xi) > 1$)  can be global. Let us  consider first flows with $\mu(\xi) > 0$.

\begin{figure}
 \epsfxsize=140mm
 \centerline{\epsfbox{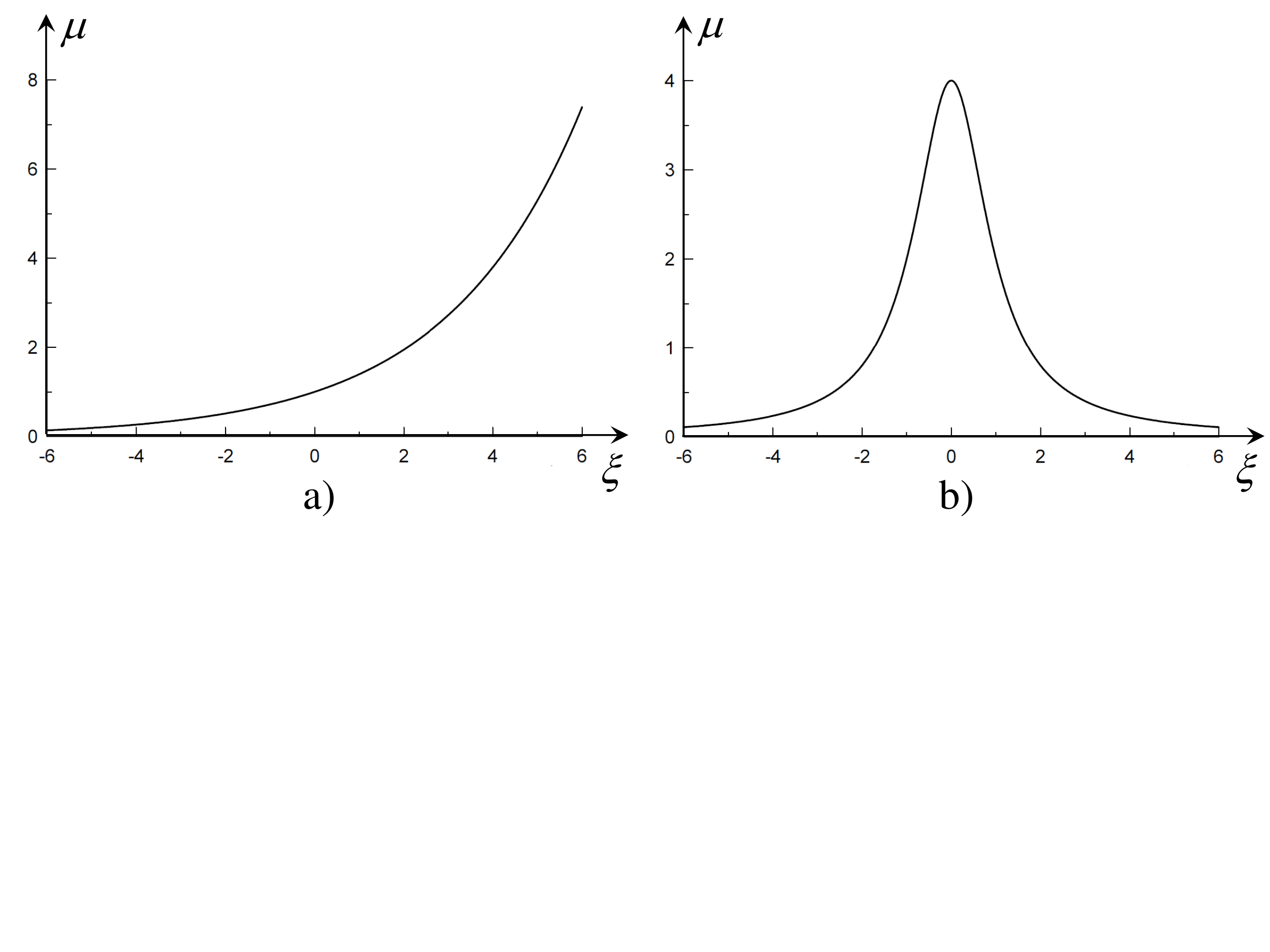}}
 \vspace{-4.5cm}
 \caption{The qualitative dependence of $\mu(\xi)$ for the flows in which $c(\xi)$ grows faster than exponentially (a) and slower than exponentially (b).}
 \label{f06}
 \end{figure}

\begin{figure}
 \epsfxsize=140mm
 \centerline{\epsfbox{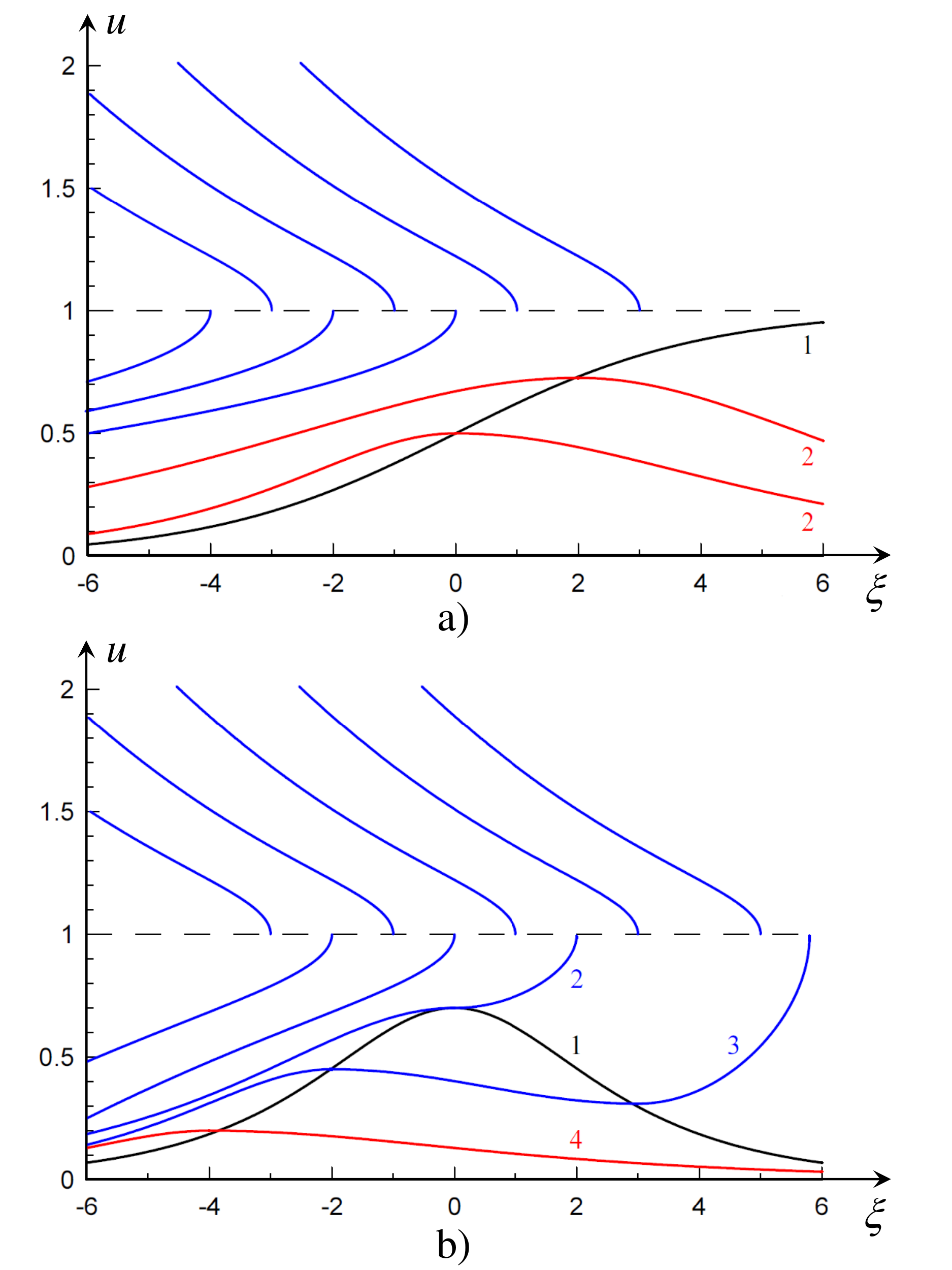}}
 %\vspace{-2cm}
 \caption{(Color online.) Solutions of equation (\ref{u1}) in the case when $c(\xi)$ grows faster than exponentially (a) and slower than exponentially (b). Line 1 depicts the null-isoclines in both frames. Lines 2 in frame (a) illustrate examples of global solutions; other non-numbered blue lines illustrate bounded solutions which exist for $\xi \le \xi_b$ with the individual value of $\xi_b$ for each line. In frame (b) one can see several non-numbered blue lines depicting bounded solutions; line 2 depicts a solution tangent to the null-isocline; line 3 is an $\xi$-bounded solution partially passing below the null-isocline; and line 4 shows a global solution.}
 \label{f07}
 \end{figure}

If function $c(\xi)$ grows faster than exponentially, then $\mu(\xi)$ monotonically increases with no limit (see figure \ref{f06}(a)). The null-isocline $u = u_0(\xi)$ also increases monotonically from zero when $\xi \to -\infty$ to unity when $\xi \to +\infty$ (see line 1 in figure \ref{f07}(a)). Solutions of equation (\ref{u1}) either intersect the null-isocline and are global, or lay entirely above it and are bounded from the right by some $\xi = \xi_b$. (Note that the null-isocline has a positive derivative, whereas solutions below it have negative derivatives. Therefore, all such solutions intersect the null-isocline when $\xi$ decreases.)
%                   Fig. 6

If function $c(\xi)$ grows slower than exponentially, then $\mu(\xi)$ has a maximum (possibly, even more than one, but it does not matter, in principle) vanishing when $\xi \to \pm \infty$ (see figure \ref{f06}(b)). The null-isocline $u = u_0(\xi)$ has qualitatively the same shape (see line 1 in figure \ref{f07}(b)). All solutions lying above or crossing its right (descending) slope (for example, like line 3 in figure \ref{f07}(b)) are, obviously, bounded from the right by some $\xi = \xi_b$. Let us consider now at what conditions global solutions such as shown by line 4 in figure \ref{f07}(b) can exist.
%                   Fig. 7

\begin{figure}
% \vspace{-.5cm}
 \epsfxsize=140mm
 \centerline{\epsfbox{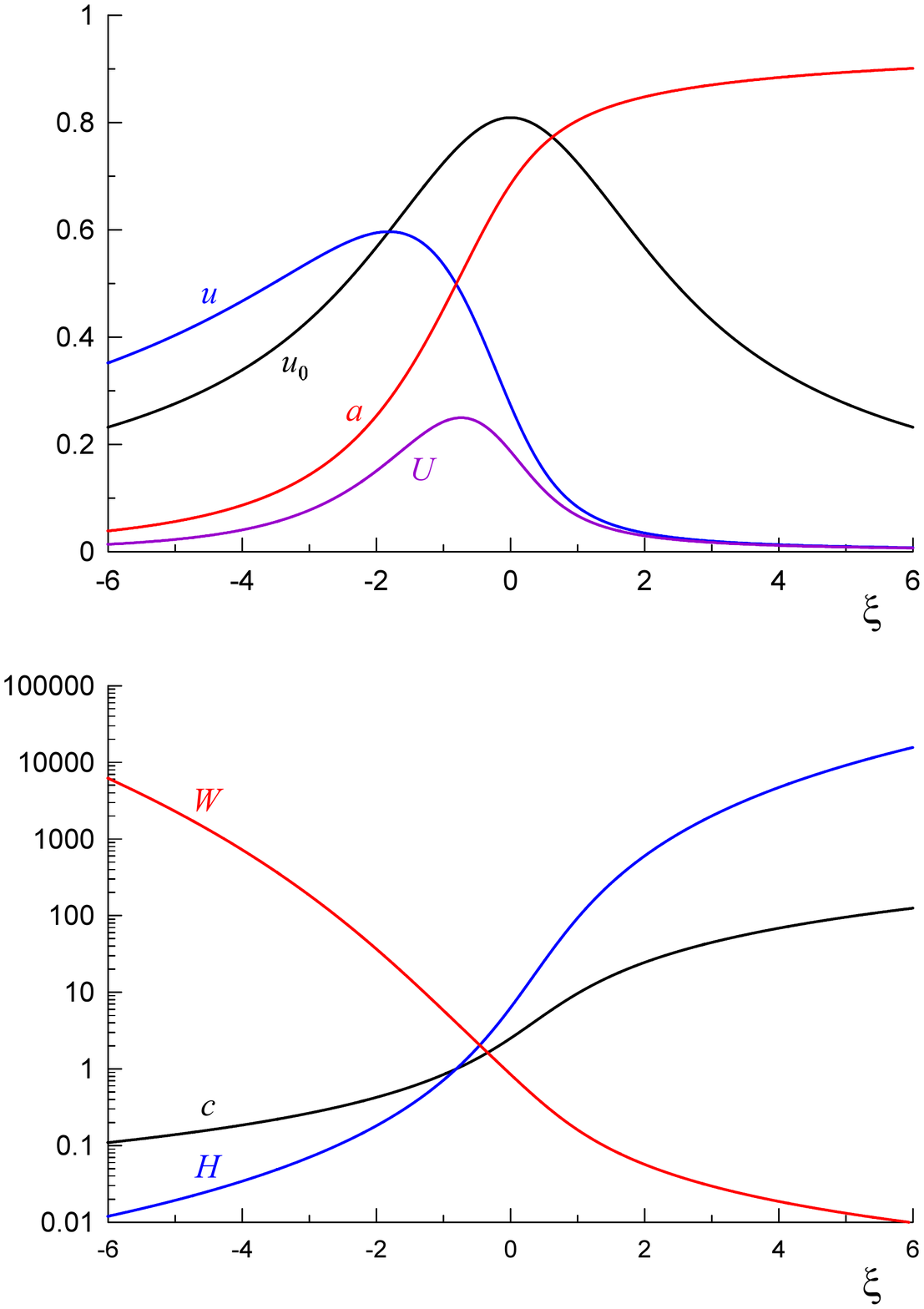}}
 \vspace{-2cm}
 \caption{(Color online.) Qualitative dependences of typical variation of parameters in the case of a subcritical global flow along the canal. All variables are normalised to some reference values and presented in dimensionless form.}
 \label{f09}
 \end{figure}

Global solutions must be below $u_0(\xi)$ for sufficiently big $\xi \ge \xi_M$. Let $\mu(\xi_M) \ll 1$, then, as follows from equation (\ref{u0}), $u_0(\xi) \approx \mu(\xi) \ll 1$ and it decreases with $\xi$ when $\xi \ge \xi_M$. Therefore, equation (\ref{u1}) simplifies and becomes:
 \[
 \dfrac{\dd u(\xi)}{\dd\xi} \approx u^2(\xi)-\mu(\xi)\,u(\xi) \equiv
 u^2(\xi) - \dfrac{u(\xi)}{c(\xi)}\,\dfrac{\dd c(\xi)}{\dd\xi}\,.
 \]
 This equation reduces to:
  \[
 \dfrac{1}{(cu)^2}\,\dfrac{\dd}{\dd\xi}(cu) = \dfrac{1}{c(\xi)}\,,
 \]
 and then, can be readily solved:
 \be
 \dfrac{1}{c(\xi)\,u(\xi)} = \dfrac{1}{c(\xi_M)\,u(\xi_M)} -
 \II{\xi_M}\dfrac{\dd y}{c(y)} + \II{\xi}\dfrac{\dd y}{c(y)}.
 \label{cu1}
 \ee
 For the convergence of the integrals, function $c(\xi)$ must grow faster than linearly, for example, as $\xi^{1 + \varepsilon}$ where $\varepsilon > 0$. Therefore, when $\xi \to +\infty$, then $c(\xi)u_0(\xi) \approx c'(\xi) \to +\infty$. Thus, the existence of global solutions is secured (i) by the convergence of the integrals in equation (\ref{cu1}) and (ii) by the inequality:
  \be
 a_0^{-1} = \dfrac{1}{c(\xi_M)\,u(\xi_M)} - \II{\xi_M}\dfrac{\dd y}{c(y)} > 0.
 \label{a0}
 \ee
 The latter condition can be easily satisfied by the choice of sufficiently small $u(\xi_M)$. 
 As the result, on the solution which passes through the point $\Bl(\xi_M,\,u(\xi_M)\Br)$, the product $c(\xi)u(\xi)$ goes to a finite limit $a_0 > 0$. 
 Due to this, $u(\xi) < u_0(\xi)$ when $\xi \ge \xi_M$, i.e. this solution is global. Thus, global subcritical flows with $D \ne 0$ and smooth profiles $c(\xi)$ and $U(\xi)$ do exist under the conditions formulated above. 
 Figure \ref{f09} demonstrates a qualitative variation of the flow parameters along the canal.
%                   Fig. 9

 To study flows with $\mu(\xi) < 0$, it is convenient to use the variable $w(\xi) = 1/u(\xi)$ (see equation (\ref{auw}) with $a(\xi) = c(\xi)/w(\xi)$). In this case, all the reasoning presented above remains almost the same. In particular, if $c(\xi)$ decreases faster than exponentially, the null-isocline and solutions on the half-plane $(\xi, \, w)$ look qualitatively the same as in figure \ref{f07}(a) (with the replace $u$ by $w$). All solutions intersecting the null-isocline are global, whereas those which are not intersecting are bounded on the right by some value $\xi_b$. If function $c(\xi)$ decreases slower than exponentially, the solutions are similar to those shown in figure \ref{f07}(b). For the existence of global solutions, it is sufficient that $c(\xi)$ decreases with $\xi$ as $\xi^{-\varepsilon - 1/3}$ when $\xi \to +\infty$ and for some point on the solution $\Bl(\xi_M, \, w (\xi_M)\Br)$, where $\xi_M \gg 1$, the following inequality holds (cf. (\ref{a0})):
 \be
 a^3(\xi_M) \equiv \left[\dfrac{c(\xi_M)}{w(\xi_M)}\right]^3  \equiv \left[c(\xi_M)u(\xi_M)\right]^3 >
 \II{\xi_M}\!c^3(y)\dd y.
 \label{a1}
 \ee
 This inequality can be easily satisfied by the choice of a sufficiently small $w(\xi_M)$ (big $u(\xi_M)$). In the conclusion, we remind that when $\mu(\xi) < 0$, all global flows are supercritical ($U>c$).
 \section{Matching solutions for reflectionless flows}
\label{Sect06}
One can show that global solutions can exist only for rather artificial velocity profiles, $c(\xi)$ and $U(\xi)$, whereas in majority of cases these profiles are defined only on the semi-axis, either on the left or on the right of some point $\xi_*$. Using solutions for $D \ne 0$ defined on the different sides of the selected common point, one can try to construct a global solution by matching two particular solutions. However, the matching procedure depends on the character of the common point, whether it is regular ($U \ne c$) or critical, where $U = c$. Below we consider both these possibilities.
 \subsection{Matching solutions in the regular point}
\label{Sect061}
 Let us assume that there are two solutions for the fluid flow satisfying equation (\ref{Eq-a}). One of them describing velocities $c_-(x)$ and $U_-(x)$ and defined for $x\le x_m$, and another solution describing velocities $c_+(x)$ and $U_+(x)$ and defined for $x\ge x_m$. Assume that $c_+(x_m)=c_-(x_m)=c_m$ and $U_+(x_m)=U_-(x_m)=U_m$ where $x=x_m$ is a regular point, i.e. such that $U_m\ne c_m$. As follows from equation (\ref{Eq-a}), transition through the point $x=x_m$ is reflectionless if $D_+ = D_-$. If $D_+ \ne D_-$, then function $a'(x)$ is discontinuous, and the coefficient $Z\sim\delta(x-x_m)$ in equation  (\ref{WEq1}), where $\delta(x)$ is the Dirac delta-function. In such a case, as will be shown in Appendix, additional waves will appear on left or right of point $x_m$. 
 
 To illustrate the possibility to construct a global solution, let us match two solutions at a regular point. To this end, we chose two solutions with equal constants $D_\pm$ obeying the equation (\ref{lb}) at different parameters $p$ and $V$. In such a case, the profiles $c(\xi)$ and $U(\xi)$ depend on the same coordinate $\xi$. Let us set
 \be
 \ba{lll}
 c_-(\xi) = c_a(\xi),\ & \ U_-(\xi) = U_a(\xi) = V_a c_a^{p_a}(\xi),\
 & \ \xi \le \xi_m; \\ \\
 c_+(\xi) = c_b(\xi),\ & \ U_+(\xi) = U_b(\xi) = V_b c_b^{p_b}(\xi),\
 & \ \xi \ge \xi_m; \\ \\
 c_a(\xi_m) = c_b(\xi_m),\ & \ U_a(\xi_m) = U_b(\xi_m). & \
 \ea
 \label{cUab}
 \ee
Then, functions $u_{a,b}(\xi) = \Bl[U_{a,b}(\xi)/c_{a,b}(\xi)\Br]^{1/2}$ satisfy equations similar to equation (\ref{u}):
 \be
 u_{a,b}^4 + 3D_{a,b}(\xi-\xi_{a,b})\,u_{a,b} + 3 = 0, \qquad
 D_{a,b}=\dfrac{p_{a,b}-1}{p_{a,b}+1}\,.
 \label{uab}
 \ee
 Functions $U_{a,b}(\xi)$ and $c_{a,b}(\xi)$ can be presented in terms of $u_{a,b}(\xi)$ using the expressions similar to equations (\ref{cUu}).  The existence of the functions $u_a(\xi)$ and $u_b(\xi)$, as well as continuity of $u(\xi)$ (i.e., $u_a(\xi_m) = u_b(\xi_m) = u_m$) are secured by the following relations:
 \be
 D_a(\xi_m - \xi_a) = D_b(\xi_m - \xi_b) < -\frac{4}{3}, \qquad \left(\dfrac{V_a}{u_m^2}\right)^{p_b-1} =
 \left(\dfrac{V_b}{u_m^2}\right)^{p_a-1}.
 \label{Vab}
 \ee

For the resulting composite flow to be global, i.e. covering the entire $\xi$-axis, the parameters $ p_ {a, b} $ should be chosen such that $D_a$ and $D_b$ have different signs, namely,
 \be
 D_a > 0\ \ \Rightarrow\ \ |p_a| > 1, \qquad
 D_b < 0\ \ \Rightarrow\ \ |p_b| < 1.
 \label{Dlb}
 \ee
It is important to remember that the value and sign of $p$ significantly impact the behavior of functions $c(\xi)$ and $U(\xi)$ when $|\xi| \to \infty$ (see equations (\ref{cU1}) and (\ref{cU2})). To find the profiles of the flow and wave velocities, in addition to $p_a$, $p_b$ and $\xi_m$, we must set, for example, the flow parameters $V_a$ and $\xi_a > \xi_m + 4/(3D_a)$ (see equation (\ref{Vab})), select from the roots of equation (\ref{uab}) the one that corresponds to the nature of the flow (subcritical or
  supercritical), and using equations (\ref{Vab}), find $V_b$ and $\xi_b$,
  \[
 V_b = \Bl[u_m^{2(p_a-p_b)}\,V_a^{p_b-1}\Br]^{\frac{1}{p_a-1}},
 \qquad \xi_b = \xi_m-\dfrac{D_a}{D_b}\Bl(\xi_m - \xi_a\Br).
 \]
 
 It should be emphasized that, with the exception of the trivial case $p_a = p_b$, the velocity profiles have a kink at $x = x_m$, i.e. their derivatives are discontinuous. Indeed, calculating with the help of (\ref{cUu}) the logarithmic derivatives of $c(\xi)$ and $U(\xi)$, we find:
 %                   Fig. 4
\begin{figure}
%\vspace{-1.0cm}
\centerline{\includegraphics[width=1.0\textwidth]{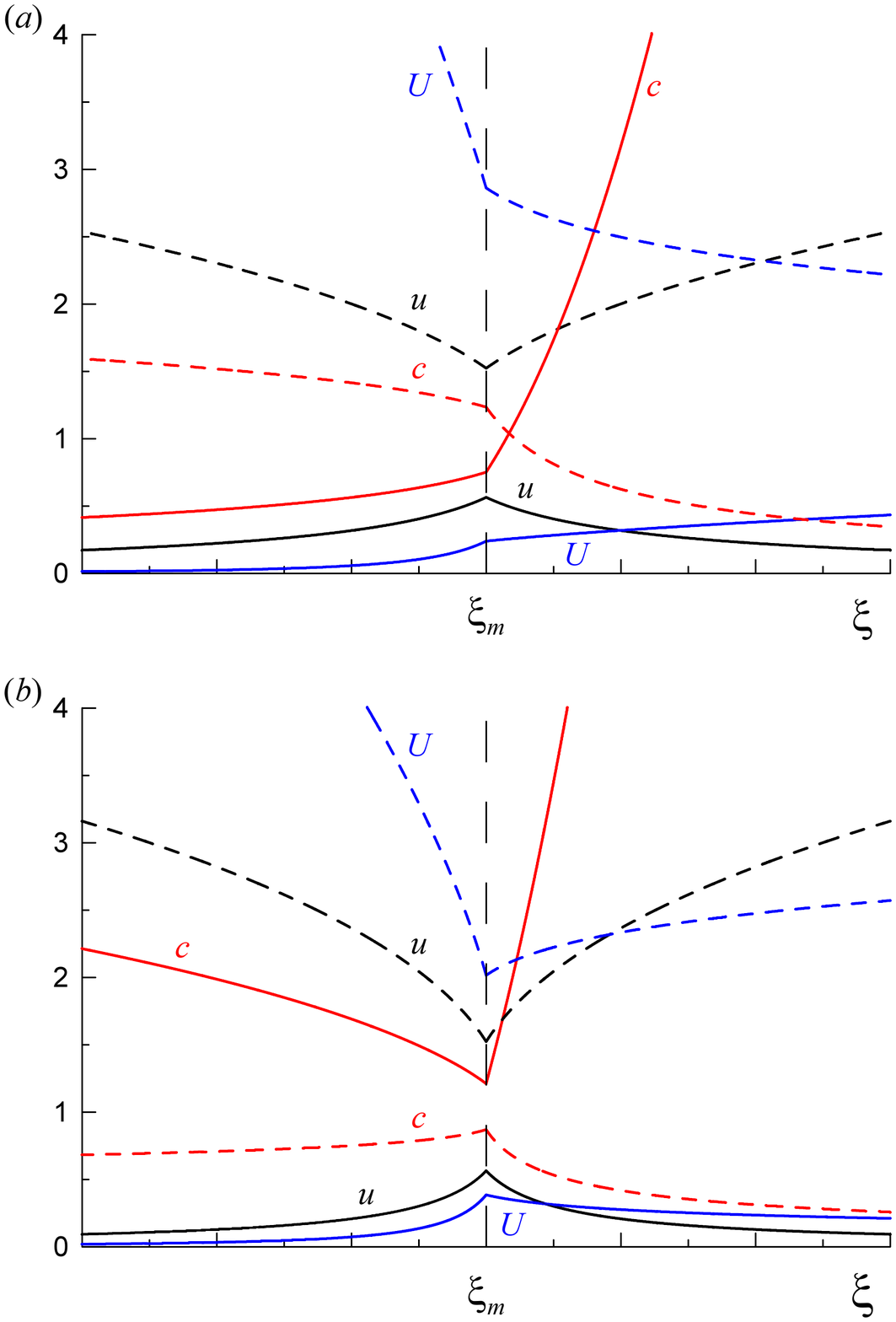}}%
\vspace{-2.cm}
\caption{The result of matching of RL flow (\ref{lb}) for different parameters:
(a) -- $p_a=5$, $V_a=1$, $p_b=1/5$; (b) -- $p_a=-5$, $V_a=1$,
 $p_b=-1/5$. Solid lines pertain to subcritical flows ($U<c$), dashed line -- 
to supercritical flows ($U>c$).}
\label{f04}
\end{figure}
\[
 \dfrac{c'_{a,b}}{c_{a,b}} = \dfrac{2}{p_{a,b}-1}\,
 \dfrac{u'_{a,b}}{u_{a,b}}, \qquad
 \dfrac{U'_{a,b}}{U_{a,b}} = \dfrac{2p_{a,b}}{p_{a,b}-1}\,
 \dfrac{u'_{a,b}}{u_{a,b}}, \qquad
 \left(\dfrac{a'(\xi)}{a(\xi)}\right)_{a,b} =
 \dfrac{p_{a,b}+1}{p_{a,b}-1}\,\dfrac{u'_{a,b}}{u_{a,b}}\,.
 \]
On the other hand, from equation (\ref{Eq-a1}) we get:
 \[
 \left(\dfrac{a'(\xi)}{a(\xi)}\right)_{a,b} =
 \dfrac{c_{a,b}^{3/2}\,U_{a,b}^{1/2}}{c_{a,b}^2-U_{a,b}^2} =
 \dfrac{u_{a,b}}{1-u_{a,b}^4}\,,
 \]
so that
\[
\left(\dfrac{u'_{a,b}}{u_{a,b}}\right)_{\xi_m} =
 \dfrac{p_{a,b}-1}{p_{a,b}+1}\,\dfrac{u_m}{1-u_m^4} \equiv
 \dfrac{p_{a,b}-1}{p_{a,b}+1}\,\dfrac{(U_m/c_m)^{1/2}}{1-(U_m/c_m)^2}\,,
\]
and then
\be
 \ba{l}
 \left(\dfrac{u'_b}{u_b} - \dfrac{u'_a}{u_a}\right)_{\xi_m} =
 \left(\dfrac{U'_b}{U_b}-\dfrac{U'_a}{U_a}\right)_{\xi_m} =
 -\left(\dfrac{c'_b}{c_b}-\dfrac{c'_a}{c_a}\right)_{\xi_m} 
 \\ \\
 \phantom{\left(\dfrac{u'_b}{u_b} - \dfrac{u'_a}{u_a}\right)_{\xi_m}} = \dfrac{2(p_b-p_a)}{(1+p_a)(1+p_b)}\,\dfrac{(U_m/c_m)^{1/2}}
 {1-(U_m/c_m)^2} \ne 0\,.
 \ea
 \label{Jump}
 \ee
 In figure \ref{f04} one can see the dependences $u$, $c$ and $U$ on $\xi$ for two versions of matching of RL flows in the regular point (the velocities $c$ and $U$ are normalised to some reference values and presented in the dimensionless form).

\subsection{Matching solutions in the critical point}
\label{Sect062}
\begin{figure}
\vspace{-3.0cm}
\centerline{\includegraphics[width=1.1\textwidth]{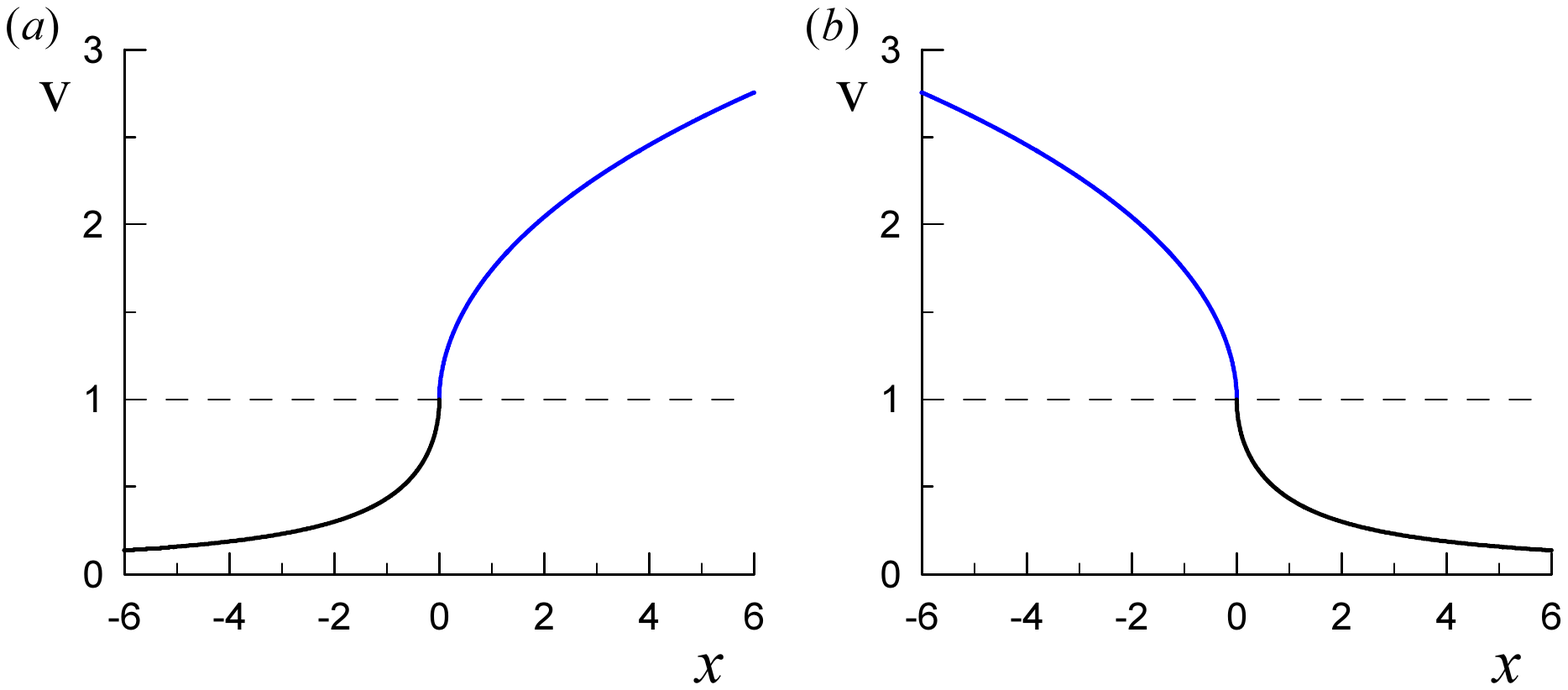}}%
\vspace{-11.5cm}
\caption{Composite solutions of equation (\ref{v}) describing the transition from the subcritical to supercritical flow (frame a) and in the opposite
 direction (frame b).}
\label{f05}
\end{figure}

Let us consider now the possibility of constructing a global solution by matching solutions with $D \ne 0$ defined on different sides of a common critical point assuming that $c(\xi) \equiv 1$. Let us set for
 certainty $\xi_0 = 4/3$ in equation (\ref{v}), then the condition of no
 reflection (\ref{Eq-a}) is satisfied only for $\xi < 0$, and a corresponding RL flow is defined only on this semi-axis. However, on the physical axis $x=\xi/D$ such a flow is defined for $x \le 0$ if $D = D^{(-)} > 0$, and for $x \ge 0$ if $D = D^{(+)} < 0$. From these solutions defined on the different semi-axes it is possible to construct a composite solution defined on the entire $x$-axis and describing the transition of the sub-critical flow into supercritical flow (see Fig. \ref{f05}a), or supercritical flow to the subcritical flow (see Fig. \ref{f05}(b). If $D^{(+)}=-D^{(-)}$, then, given the asymptotic expansions (\ref{xi-0}), the velocity profile $U(x)$ looks quite ``smooth'' (if the infinite derivative at $x=0$ is ignored). 
 The theory of wave propagation on shallow water flows is well-developed including the cases when the current speed passes through the critical point having {\it finite derivatives} of the velocity profiles $U(x)$ and $c(x)$ \citep[see, for example][and references therein]{ChEst-17, ChSt21}. 
 However, in the discussed case, a question about the contribution of the {\it singular} (sic!) point $x=0$ to the possible generation of additional waves still remains open. 
 Potentially some other waves than the incident and transmitted waves can emerge from the critical point. 
 In addition, at least partial absorption of an incident wave can occur in the critical point.

 In a similar way, one can construct a matched solution for a flow with a constant velocity with the only difference being that in this case the solution on the left ($x\le 0$) must have $D=D^{(-)} < 0$, whereas the solution on the right ($x\ge 0$) must have $D=D^{(+)} > 0$.
%                   Fig. 5
 \section{Conclusion}
\label{Sect07}
In this paper, we have shown that reflectionless (RL) propagation of long surface waves in a duct or in a channel with the variable width $W(x)$ and depth $H(x)$ in the presence of a spatially varying flow is quite possible if the velocities of the flow $U(x)$ and waves  $c(x) = \sqrt{gH(x)}$ are interrelated by equation (\ref{Eq-aa}). It is important to emphasize that this single equation cannot define both functions $c(x)$ and $U(x)$ simultaneously. Therefore, it is necessary either to define one of them or set some additional relation between them. In the result, we arrive at the two infinite classes of RL flows distinguishing in that whether or not the constant $D$ in equation (\ref{Eq-aa}) is equal to zero.

If $D = 0$, the product $c(x)U(x)$ is constant along the canal (see condition (\ref{M1})).  All RL flows of this class have an interesting feature, the wider is the canal, the higher is the flow speed $U(x)$. Indeed, condition (\ref{M1}) is equivalent to $U(x) H^{1/2}(x) = \mbox{const}$. Then, from the flux conservation equation (\ref{Flux}) it follows that $W(x) H^{1/2}(x) = \mbox{const}$, so that $U(x) \sim W(x)$. Note that the same relation between the water depth and duct width was obtained by  \citet{Pelinovsky-17} to describe global RL solutions for the case when there is no current.

This class of RL flows has an advantage that we can attribute arbitrarily the $x$-dependence to any of two velocities $c(x)$ or $U(x)$, as well as the value to their product. Doing so, we can also attribute to $c(x)$ and $U(x)$ such important properties as continuity, differentiability,  boundedness, and so on. Moreover, we can define the flow on the entire $x$-axis, i.e. globally. As a result, the considered class contains currents with a wide variety of properties, including subcritical ($U < c$) and supercritical ($U > c$) flows, flows with transition from the subcritical to the supercritical regime and/or vice versa, as well as currents with other characteristics.

In contrast, when $D \ne 0$, the properties of RL flows can vary in a much more restricted range. First of all, the currents of this class can be either subcritical, or supercritical, because a transition through the critical point (where $U = c$)  most likely violates the RL property of the flow. Respectively, the majority of RL flows are defined only on a certain interval of the $x$-axis, namely, on a ray bounded by the critical point $x_*$ at which $c(x_*) = U(x_*)$ (see Section \ref{Sect04}). Solutions are defined either on the right or on the left of $x_*$, depending on the parameters of the problem, e.g., on the sign of $D$.

In global flows of this class, as shown in Section \ref{Sect05}, the  profiles of $c(x)$ and $U (x)$ can be smooth functions of $x$ only if one of them, say $c(x)$, is varying (increasing or decreasing) fast enough. Global RL flows can also be constructed by matching bounded flows with equal values of $D$ and overlapping domains of definition in some regular point (where $c \ne U$) common to both flows. However, at this point the profiles of $c(x)$ and $U(x)$ have necessarily kinks (jumps in their derivatives -- see Section \ref{Sect061}). With this in mind, one can construct a piece-wise smooth flow by matching solutions many times in different points. Matching the currents with different values of $D$ is also possible, but this will inevitably lead to the occurrence of additional waves, i.e. to the loss of the RL property of the flow (see Appendix).

The importance of the RL flows is that long surface waves of arbitrary form in the linear approximation can independently propagate in the opposite directions of the $x$-axis. This provides the most efficient energy transmission over a long distance through the inhomogeneous environment. The results obtained in this paper generalise earlier derived results \citep[see][]{Choi-08,  Did-Pel-Soom-09, Did-Pel-09, Didenkulova-09, Grimshaw-10, Did-Pel-11} for the cases when a water flow plays an important role and must be taken into consideration. However, we did not consider here two- or three-dimensional effects which were studied to a degree in some of the cited papers. This interesting issue can be investigated elsewhere later. The results obtained can be of interest to mitigate the possible impact of waves on ships and marine engineering constructions in the coastal zones. \\

{\bf Acknowledgements.} S.C. was financially supported by the Ministry of Science and Higher Education of the Russian Federation.
 Y.S. acknowledges the funding of this study provided by the grant No. FSWE-2020-0007 through the State task program in the sphere of scientific activity of the Ministry of Science and Higher Education of the Russian Federation, and the grant No. NSH-2485.2020.5 provided by the President of Russian Federation for the State support of leading Scientific Schools of the Russian Federation. 

%%%%%%%%%%%%%%%%%%%%%%%%%%%%%%%%%%%%%%%%%%%%%%%%%%%%%%%%%%%%%%%%%%%%%%%
\appendix
\section{Wave scattering in the matching point}
\label{appA}

In this Appendix, we show that if $D_+ \ne D_-$, then a composite flow will not be reflectionless. The ensemble of waves in the vicinity of the regular matching point is determined by whether the matching flows are subcritical ($U_m<c_m$) or supercritical ($U_m>c_m$). Solution of Eq. (\ref{WEq1}) obviously is continuous, $\psi_+(x_m,t) = \psi_-(x_m,t) \equiv \psi_m(t)$, and its time derivatives are continuous too. By integrating equation (\ref{WEq1}) over $x$ within the interval from $x_m-\vep$ to $x_m+\vep$ and denoting $[f] = f(x_m+\vep)-f(x_m-\vep)$ at $\vep\to +0$, we obtain:
\[
 \left[\dfrac{\pl\psi}{\pl x}\right] +
 \dfrac{1}{a_m}\left[\dfrac{\dd a}{\dd x}\right]\psi_m(t) = 0, \qquad
 a_m = a(x_m) \equiv \Bl(c_mU_m\Br)^{1/2}.
 \]
Using Eq. (\ref{Eq-a}), we arrive at the matching conditions:
 \be
  \Bl[\psi\Br] = 0, \qquad
 \left[\dfrac{\pl\psi}{\pl x}\right] + \dfrac{D_+-D_-}{c_m^2-U_m^2}
 \,a_m c_m\psi_m(t) = 0.
 \label{Match}
 \ee 

 {\bf i)} Let us assume that the flows are subcritical in the vicinity of point $x_m$, ($U_m < c_m$), then solution to the left of point $x_m$ can contain the incident and reflected waves, whereas solution to the right of this point contains only a transmitted wave (see Eq. (\ref{psi})):
 \[
 \ba{l}
 \psi_-(x,t) = \psi_1\left(t-\din_{x_m}^{x}\dfrac{\dd z}
 {U_-(z)+c_-(z)}\right)
 + \psi_2\left(t+\din_{x_m}^{x}\dfrac{\dd z}{c_-(z)-U_-(z)}\right),
     \\ \\
 \psi_+(x,t) = \psi_3\left(t-\din_{x_m}^{x}\dfrac{\dd z}
 {U_+(z)+c_+(z)}\right).
 \ea
 \]
 Matching these solutions in accordance with the conditions (\ref{Match}) leads to the equation:
 \[
 \psi'_2(t) = \mu\Bl(\psi_1(t)+\psi_2(t)\Br),\qquad
 \mu = \dfrac{a_m}{2}\Bl(D_+ - D_-\Br).
 \]
 Solving this equation under assumption that function $\psi_1(t)$ vanishes sufficiently quickly when $t \to -\infty$, we derive:
 \be
 \psi_2(t) = \mu\!\din_{-\infty}^{t}\!\psi_1(\tau)\re^{\mu(t-\tau)}\,\dd\tau,
 \qquad \psi_3(t)=\psi_1(t) + \psi_2(t).
 \label{sub}
 \ee
 It is easy to see that $\psi_2(t) \equiv 0$ only if $\psi_1(t) \equiv 0$.\\

 {\bf ii)} For the matching of supercritical flows ($U_m > c_m$), we need to consider separately two cases.
 
 If a wave of a positive energy arrives from the left to the point $x_m$, then we have:
\[
 \ba{l}
 \psi_-(x,t) = \psi_1\left(t-\din_{x_m}^{x}\dfrac{\dd z}
 {U_-(z)+c_-(z)}\right)\,,
      \\ \\
 \psi_+(x,t) = \psi_3\left(t-\!\din_{x_m}^{x}\!\dfrac{\dd z}
 {U_+(z)+c_+(z)}\right) +
 \psi_4\left(t-\!\din_{x_m}^{x}\!\dfrac{\dd z}{U_+(z)-c_+(z)}\right),
 \ea
 \]
 Matching solutions in the point $x_m$ leads to the equations $\psi'_4(t)+\mu\psi_1(t)=0$, so that we obtain:
 \be
 \psi_4(t) = -\mu\!\!\din_{-\infty}^{t}\!\psi_1(\tau)\,\dd\tau, \qquad \psi_3(t) = \psi_1(t) - \psi_4(t).
 \label{super-p}
 \ee
 
 If a wave of a {\it negative energy} \citep{Fabr-Step} arrives from the left to the point $x_m$, then we have:
 \[
 \ba{l}
 \psi_-(x,t) = \psi_2\left(t-\din_{x_m}^{x}\dfrac{\dd z}
 {U_-(z)-c_-(z)}\right)\,,
      \\ \\
 \psi_+(x,t) = \psi_3\left(t-\din_{x_m}^{x}\dfrac{\dd z}
 {U_+(z)+c_+(z)}\right) +
 \psi_4\left(t-\din_{x_m}^{x}\dfrac{\dd z}{U_+(z)-c_+(z)}\right),
 \ea
 \]
  Matching of these solutions leads to the equation $\psi'_3(t) = \mu\,\psi_2(t)$, so that we obtain:
 \be
 \psi_3(t) = \mu\!\!\din_{-\infty}^{t}\!\psi_2(\tau)\,\dd\tau, \qquad \psi_4(t) = \psi_2(t) - \psi_3(t).
 \label{super-n}
 \ee
Let us note in passing that if a superposition of waves with the positive and negative energies arrives from the left to the point $x_m$, $\psi_- = \psi_1 + \psi_2$, then
 \begin{eqnarray}
 \psi_3(t) &=& \psi_1(t) + \mu\!\!\din_{-\infty}^{t}\!\Bl(\psi_1(\tau) +
 \psi_2(\tau)\Br)\,\dd\tau, \label{superA}\\
 \psi_4(t) &=& \psi_2(t) - \mu\!\!\din_{-\infty}^{t}\!\Bl(\psi_1(\tau) +
 \psi_2(\tau)\Br)\,\dd\tau. \label{superB}
 \end{eqnarray}

Thus, we have seen that when $D_+ \ne D_-$, the composite flow can not be a RL flow in any regime, subcritical or supercritical.
%%%%%%%%%%%%%%%%%%%%%%%%%%%%%%%%%%%%%%%%%%%%%%%%%%%%%%%%%%%%%%%%%%%%%%%
\bibliographystyle{jfm}
\bibliography{Chur-Step}

\end{document}